\documentclass[draft]{beatcs}
\usepackage{amsmath,amssymb,amsfonts}
\title{On  ground word problem of term equation systems}
\author{S\'andor V\'agv\"olgyi \\
        Department  of Foundations of Computer Science\\
                 University of Szeged\\
        H-6720 Szeged, \'Arp\'ad t\'er 2, Hungary\\
          Email: vagvolgy@inf.u-szeged.hu
}

\let\s=\sigma         \let\b=\Box
\let\S=\Sigma

\def\ts{T_\Sigma}

\def\les#1#2#3{#1\leq #2\leq #3}

\def\seq#1#2#3{#1_{#2},\ldots,#1_{#3}}

\def\red#1{\mathop\rightarrow_{#1}}
\def\tred#1{\mathop\rightarrow_{#1}^{*}}
\def\thue#1{\mathop\leftrightarrow_{#1}}
\def\tthue#1{\mathop\leftrightarrow_{#1}^{*}}

\def\pr{{\bf Proof.\ }}

\def\-{$-$}

\newtheorem{tet}{Theorem}[section]
\newtheorem{sta}[tet]{Statement}

\newtheorem{cla}[tet]{Claim}
\newtheorem{prop}[tet]{Proposition}
\newtheorem{lem}[tet]{Lemma}
\newtheorem{df}[tet]{Definition}
\newtheorem{exa}[tet]{Example}

\newtheorem{ques}[tet]{Open Question}

\begin{document}
\date{}
\maketitle
\begin{abstract}
We give    semi-decision procedures for the ground
             word problem of variable preserving term equation systems and 
term equation systems. 
They are natural improvements of two  well known trivial  semi-decision procedures. We show the correctness of our procedures.
\end{abstract}

Keywords: term equation systems; ground word problem; 
Knuth-Bendix  
completion procedure; 
ground term rewriting systems
\section{Introduction}

A term equation  $l \approx r$ is called  
variable preserving if the same variables  occur in the left-hand side 
$l$ as in  the right-hand side $r$. 
 A term equation system (TES) $E$ is  called variable preserving
 if all of its equations are variable preserving.
The ground word problem is undecidable even for variable-preserving TESs, see Example 4.1.4 on page 60 in  \cite{baanip}. 
We recall the well known trivial  semi-decision procedure $\textit{PRO1}$
 for the ground word problem of  variable preserving TESs 
and its straightforward generalization, 
the  trivial  semi-decision procedure $\textit{PRO2}$  for the ground word problem of TESs. 

On the basis of $\textit{PRO1}$, we  give a    semi-decision procedure $\textit{PRO3}$ for the ground
 word problem of   variable preserving TESs. 
Given a  TES $E$  and ground terms $p, q$ over the ranked alphabet $\S$,
procedure $\textit{PRO3}$ constructs the ground TESs
(GTESs) $P_i$ and  $Q_i$, $i\geq 1$ such that

$(a) $ $P_i \cup Q_i \subseteq \tthue E$ for $i\geq 1$. 

\noindent
Condition (a) ensures that the congruence closure of $P_i \cup Q_i$ is a subset of $\tthue E$. 

\noindent  Procedure $\textit{PRO3}$ outputs an answer and halts if and only if
 
$(b)$ there is a $j\geq 1$ such that

\noindent 
$p\tthue {P_j\cup Q_j} q$ or 

\noindent 
$\tthue {P_j}\cap (\{\,  p\, \}\times \ts)=\tthue E\cap (\{\,  p\, \}\times \ts) $ or 

\noindent
$\tthue {Q_j}\cap (\{\,  q\, \}\times \ts)=\tthue E\cap (\{\,  q\, \}\times \ts) $.

\noindent
Condition (b) says that we have a proof of $ p\tthue E q$,  or
the intersection of $\tthue {P_j}$ with  $ (\{\,  p\, \}\times \ts)$ is equal to that of  $\tthue E$,  
or the intersection of  $\tthue {Q_j}$ with  $(\{\,  q\, \}\times \ts) $ is equal to that of  $\tthue E$. 
Assume that (b) holds. 
If
$p \tthue {P_j \cup Q_j} q$ holds, $\textit{PRO3}$ outputs 'yes', and halts.
Otherwise, 
if 

$\bullet$ 
the intersection of $\tthue {P_j}$ with $ (\{\,  p\, \}\times \ts)$ is equal to that of  $\tthue E$,  
or

$\bullet$ 
 the intersection of  $\tthue {Q_j}$ with  $(\{\,  q\, \}\times \ts) $ is equal to that of  $\tthue E$,

\noindent
then $p\tthue E q$ does not hold either. Hence  semi-decision procedure $\textit{PRO3}$ outputs
'no' and halts.

Procedure $\textit{PRO3}$ constructs the ground TESs
(GTESs) $P_i$ and  $Q_i$, $i\geq 1$ in the following way.
We put a ground instance  $l'\approx r'$ of an equation  $l\approx r$ of $E\cup E^{-1}$ in $P_1$ 
if $l'$ is a subterm of $p$. Then we 
iterate the following computation items.

$\bullet$ We convert the GTES $P_i$ into an equivalent   reduced ground term rewrite system $R_i$
applying Snyder's fast ground completion algorithm \cite{sny}.

$\bullet$ We define the GTES $P_{i+1}$ from the 
 reduced ground term rewrite system
$R_i$ by adding  all ground instances $l\approx r$ of equations in  $E\cup E^{-1}$
such that  

- $l\approx r$ is not in $\tthue {P_i}$ and that 

- there exists  a term $s$ such that  the conversion  $p\tthue {P_i} s $ 
can be continued applying $l\approx r$ to $s$. If $P_{i+1}=R_i$, then we let $R_{i+1}=R_i$, and hence  $R_i=P_j=R_j$ holds for  $j\geq i+1$. 

\noindent Here we consider both the 
 reduced ground term rewrite system
$R_i$ and  the GTES $P_{i+1}$ as  subsets of $\ts \times \ts$. 
Furthermore, we consider a ground instance of an equation  in  $E\cup E^{-1}$
as an element of 
$\ts \times \ts$. 

  We define  the GTES $Q_i$ symmetrically to $P_i$ for $i \geq 1$. 

Procedure $\textit{PRO3}$ computes in the following way. For each $i=1, 2, \ldots$, 

$\bullet$ if $p\tthue {P_i\cup Q_i} q$, then we  output the answer 'yes'  and  halt;

$\bullet$ otherwise, if $i\geq 2$ and we did not add ground instances of equations in  $E\cup E^{-1}$
 to  the reduced ground term rewrite system $R_{P_{i-1}}$, equivalent to $P_{i-1}$,
or  to  the reduced ground term rewrite system $R_{Q_i}$, equivalent to $Q_{i-1}$,
 in the previous iteration step, 
then    we  output the answer 'no'  and  halt.  

Assume that $p\tthue E q$. Then, at some step  during the run of procedure $\textit{PRO3}$,
 $p\tthue {P\cup Q} q$ becomes true,  and procedure $\textit{PRO3}$ outputs 'yes' and halt.
If  $p\tthue E q$ does not hold, then procedure  $\textit{PRO3}$ either outputs 'no' and halts or  runs forever.

We  give a    semi-decision procedure $\textit{PRO4}$ for the ground  word problem of TESs. We obtain it generalizing  $\textit{PRO3}$   taking into account
$\textit{PRO2}$. The main difference is the following.
 We define $P_{i+1}$ from $R_i$ by adding all ground instances 
$l'\approx r'$ of the equations  $l\approx r$  in  $E\cup E^{-1}$
such that  

$\bullet$ 
$l'\approx r'$ is not in $\tthue {P_i}$, that

$\bullet$  there exists  a term $s$ such that
 a conversion  $p\tthue {P_i} s $ can be continued applying 
$l\approx r$ to $s$, and that

$\bullet$ 
we substitute some finitely many  ground terms depending on $i$, $R_i$, and $p$, for those  variables in $r$ that do not appear in $l$.

\noindent
We modify the halting condition of the proceedure so that it stops 
 if we did not add ground instances of equations in  $E\cup E^{-1}$
 to  $P_i$ or $Q_i$ in two successive  iteration steps.
 We need two successive steps rather than one. Because, in general, the heights  of the 
substituted terms becomes larger in each step. If we  do not add ground  equations to  $P_i$ in a step, then in the next step we still may add ground equations  to  $P_i$.

Procedures   $\textit{PRO3}$ and $\textit{PRO4}$ compute in a different way than all versions of the 
Knuth-Bendix completion procedure. 
To some instances of   the ground word problem of a TES $E$,
procedures   $\textit{PRO3}$ and $\textit{PRO4}$ give an answer 
sooner than all versions of the  Knuth-Bendix completion procedure
or it is open whether some  version of the  Knuth-Bendix completion procedure gives an answer at all.
Consequently, they  may  compute efficiently for some instances of   the ground word problem of a TES $E$, when the various versions of the 
Knuth-Bendix completion procedure does not give an answer  to the ground word problem of a TES $E$ 
at all or at least not in a reasonable time. However, it is still open in which cases are $\textit{PRO3}$ and $\textit{PRO4}$ really efficient.

In Section \ref{z1},  we present a brief review of the notions, notations,  and 
preliminary results used in the paper. In Section \ref{rur} we introduce and study the concept of reading-up reachability 
for reduced ground term rewriting systems.
In Section \ref{led} we present the procedures $\textit{PRO1}$ and $\textit{PRO2}$. 
In Section \ref{fo},  we present the procedure $\textit{PRO3}$, and show its correctness.  We give examples when procedure $\textit{PRO3}$ is more efficient  than procedure $\textit{PRO1}$.
In Section \ref{2fo},   we present the procedure
$\textit{PRO4}$,  and show its correctness.
In Section \ref{foglal},  we 
compare  procedures $\textit{PRO3}$ and $\textit{PRO4}$ with the basic  Knuth-Bendix completion procedure 
(see Section 7.1 in \cite{baanip}), an improved version of the  Knuth-Bendix completion procedure
 described by a set of inference rules 
(see Section 7.2  in \cite{baanip}),
the  goal-directed completion procedure based on SOUR graphs  \cite{lyn,ls}, and the 
unfailing  Knuth-Bendix completion procedure \cite{bdp}. 
In Section \ref{osszefog},  we sum up our results, and explain the applicability of 
procedures $\textit{PRO3}$ and $\textit{PRO4}$.

\section {Preliminaries}\label{z1}

In this section we present a brief review of the notions, notations and
preliminary results used in the paper. For all unexplained notions and notation see \cite{baanip}.

{\bf Relations.}
Let $\rho$ be an
equivalence relation on $A$. Then for
every $a\in A$, we denote by $a/\rho$
the $\rho$-class containing $a$,
i.e. $a/\rho=\{\,b\mid a\rho b\,\}$.
For each $B\subseteq A$, let $B/\rho=\{\,  b/\rho \mid b\in B\, \}$.

\subsection{Abstract Reduction Systems}
An abstract reduction system is a pair $(A, \rightarrow)$, 
where the reduction $\rightarrow$ is a binary relation on the set $A$. 
 $\rightarrow^{-1}$,  $\leftrightarrow $, $\rightarrow^*$, and $\leftrightarrow^* $ denote 
the inverse, the symmetric closure,  the reflexive transitive closure, and the reflexive transitive symmetric closure
of  the  binary relation $\rightarrow$, respectively.

$\bullet$
$x\in A$ is reducible if there is  $y$ such that $x\rightarrow y$. 

$\bullet$
$x\in A$ is irreducible if it is not reducible.

$\bullet$
$y\in A$ is a normal form of $x\in A$ if
$x\rightarrow ^* y$ and $y$ is irreducible. 
 If $x\in A$ has a unique normal form, the latter is denoted by  $x\hspace{-1mm}\downarrow$.

$\bullet$ $y\in A$ is a descendant of $x\in A$ if $x \rightarrow ^* y$. 

$\bullet$
$x\in A$ and $y\in A$ are joinable if there is a $z$ such that $x\rightarrow ^*z \leftarrow ^* y$, in which case we write 
$x \hspace{-1mm}\downarrow \hspace{-1mm} y$. 

\noindent
The reduction 
$\rightarrow$ is called

$\bullet$ 
 confluent if for all $x, y_1, y_2\in A$, if $y_1 \leftarrow ^*   x\rightarrow ^* y_2$, then 
$y_1 \hspace{-1mm}\downarrow \hspace{-1mm} y_2$;

$\bullet$ 
 locally confluent if for all $x, y_1, y_2\in A$, if $y_1 \leftarrow    x\rightarrow  y_2$, then 
$y_1 \hspace{-1mm}\downarrow \hspace{-1mm} y_2$;

$\bullet$ 
terminating if there is no infinite chain 
$x_0\rightarrow x_1 \rightarrow x_2 \rightarrow \cdots$;

$\bullet$ convergent if it is both confluent and terminating.

\noindent
If  $\rightarrow$ is convergent, 
then  each $x\in A$ has a unique normal form  \cite{baanip}.

{\bf Terms.}  
A ranked alphabet $\S$ is a finite set of symbols in
which every element has a unique rank in the set of nonnegative integers. For each integer $m\geq 0$, $\S_m$
denotes the elements of $\S$ which have rank $m$. 

Let $Y$ be a set of variables. The set of terms over $\S$ with variables in 
$Y$ is denoted by $\ts(Y)$. 
The set $T_{\S}(\emptyset)$ is written simply as $T_\S$ and called the set of
ground terms over $\S$.
We specify a  countably infinite set
$X=\{\,x_1,x_2,\ldots\,\}$ of variables which will be kept fixed in this paper.
Moreover, we put  $X_n=\{\,x_1, x_2, \ldots, x_n\,\}$, for $n\geq 0$. Hence $X_0=\emptyset$. 
For any $i\geq 1$ and $ j\geq 0$, let $X_{[i, j]}= \emptyset$ if $i>j$, and 
let $X_{[i, j]}= \{\, x_i, x_{i+1},\ldots, x_{j}\, \}$ otherwise. 

 For a term $t\in \ts (X)$, the height
$height(t)\in N$ 
is defined by recursion:

(a) if $t\in \S_0\cup X$, then 
$height(t)=0$, 

(b) if $t=\s(\seq t1m)$ with $m\geq 1$ and $\s \in \S_m$, then

\hspace{1 cm}$height(t)=1+max(height(t_i)\mid 1\leq i\leq m )$.

\noindent 
For each $k\geq 0$, 
$HE_{\S, \leq k}(X)=\{\, t\in \ts(X)\mid height(t) \leq k \, \}$.

Let $N$ be the set of all positive integers.
$ N^*$ stands for the free monoid generated by $N$ with empty word 
$\lambda$ as identity element. For each word $\alpha \in  N^*$,
$length(\alpha)$ stands for the length of $\alpha$.
Consider the words $\alpha, \beta, \gamma \in 
 N^*$ such that $\alpha= \beta\gamma$. Then we say that $\beta$ is a prefix of $\alpha$. Furthermore, if $\alpha\neq \beta$, then $\beta$ is a proper prefix of $\alpha$.
For a term $t\in \ts (X)$, the set
$Pos(t)\subseteq N^*$  of positions  is defined by recursion:

\begin{itemize}
\item[(i)] if $t\in \S_0\cup X$, then
$Pos(t)=\{\,\lambda\,\}$, and
\item[(ii)] if $t=\s(\seq t1m)$ with $m\geq 1$ and $\s \in \S_m$, then
$Pos(t)=\{\,\lambda\,\}\cup\{\,i\alpha\mid 1\leq i\leq m$ and $\alpha\in Pos(t_i)\,\}$.
\end{itemize}

For each term $t\in \ts (X)$, $size(t)$ is the cardinality of $Pos(t)$.

For each $t\in T_\S(X)$ and $\alpha\in Pos(t)$, we introduce the subterm 
$t/\alpha\in \ts(X)$ of $t$ at $\alpha$ 
as follows:
\begin{itemize}
\item[(a)] for $t\in \S_0\cup X$, $t/\lambda =t$;

\item[(b)] for $t=\s(\seq t1m)$ with $m\geq 1$ and $f\in \S_m$, if 
$\alpha =\lambda$ then $t/\alpha=t$,

 otherwise, if
 $\alpha =i\beta$ with $1\leq i\leq m$, then $t/\alpha=t_i/\beta$ 
\end{itemize}

For any $t\in\ts(X)$, $\alpha\in Pos(t)$, and $r\in\ts(X)$, we define
$t[\alpha\leftarrow r]\in\ts(X)$.

\begin{itemize}
\item[(i)] If $\alpha=\lambda$, then $t[\alpha\leftarrow r]=r$.

\item[(ii)] If $\alpha=i\beta$, for some integer $i$, then $t=\s(\seq t1m)$ with $f\in\S_m$ and $\les 1im$. Then $t[\alpha\leftarrow r]
=\s(t_1,\ldots,t_{i-1},t_i[\beta\leftarrow r],t_{i+1},\ldots,t_m)$.
\end{itemize}

For a term $t\in \ts(X)$, the set $sub(t)$ of subterms
of $t$ is defined as $sub(t)= \{\, t /\alpha \mid \alpha \in Pos(t) \, \}$.


 Given a term $t\in T_\S(X_n)$, $n\geq 0$,  and terms
$t_1,\ldots,t_n$, we denote by $t[t_1,\ldots ,t_n]$ the term
which can be obtained from $t$ by replacing each occurrence of $x_i$ in $t$ by $t_i$ for $1\leq i\leq n$.
A context is a  term $u \in T_{\S \cup \{\, \diamond\, \}}$, where the nullary symbol $\diamond$ appears exactly once in $u$.  
We denote the set of all contexts over $\S$ by $C_\S$. 
For a context $u$ and a term $t$,  $u[t]$ is defined from $u$
by replacing the  occurrence of  $\diamond$ with $t$.

For the sake of simplicity,  we may write  unary terms  as strings. For example, we write $fgh\#$ for the term $f(g(h(\#)))$
 and $f^3 x_1$ for $f(f(f(x_1)))$,  where
$f, g, h$ are unary symbols and $\#$ is a nullary symbol.

{\bf Algebras.} Let $\S$ be a ranked alphabet. A
$\S$ algebra is a system ${\bf B}=(B,\S^{\bf B})$, where $B$ is a nonempty set, called the carrier set of ${\bf B}$, and
$\S^{\bf B}=\{\,f^{\bf B}\mid f\in\S\,\}$ is a
$\Sigma$-indexed family of 
operations over $B$ such that for every $f\in\S_m$ with $m\geq 0$, $f^{\bf B}$ is a mapping from $B^m$ to $B$.
An equivalence relation $\rho\subseteq B\times B$ is a congruence
on ${\bf B}$ if $$f^{\bf B}(t_1, \ldots, t_m)\rho
f^{\bf B}(p_1,\dots, p_m)$$ whenever $f\in\S_m$, $m\geq 0$, and
$t_i\rho p_i$, for $1\leq i\leq m$. For each $B'\subseteq B$, let
 $[B']_\rho=\{\, [b]_\rho \mid b \in B'\, \}$.
In this paper we shall mainly deal with the algebra ${\bf
TA}=(\ts,\S)$ of ground terms over $\S$, where for any $f\in \S_m$ with
$m\geq 0$ and $t_1,\ldots ,t_m\in \ts$, we have $$f^{\bf
TA}(t_1,\ldots ,t_m)=f(t_1,\ldots ,t_m)\,.$$
We now recall the concept of a set of representatives for a
congruence $\rho$ and a set of $\rho$-classes.
\begin{df}\label{yyy}{\em \cite{fulvagacta} 
Let $\rho$ be a  congruence on ${\bf TA}$ and let $A$ be a
set of $\rho$-classes. A set $REP$ of ground terms is called a set of
{\em representatives} for $A$ if
\begin{itemize}
\item $REP\subseteq \bigcup A$,
\item $\bigcup(\, sub(t)\mid t\in REP\,)\subseteq REP $, and
\item each class $Z\in A$ contains exactly one term
$t\in REP$.
\end{itemize}}
\end{df}

{\bf Term equation systems.}
Let $\S$ be a ranked alphabet. A term equation system (TES for short)
$E$ over $\S$ is a finite
subset of $\ts(X)\times\ts(X)$.  Elements $(l,r)$ of $E$ are 
called equations and are denoted
by $l\approx r$. The reduction relation $\red E \subseteq \ts(X) \times \ts(X)$ is defined as 
follows. 
For any  terms $s, t\in \ts(X)$, $s \red E t$ if there is a pair $l\approx r$
in $E$ and 
a context $u\in C_\S(X_1)$ and a substition $\delta$ such that $s=u[\delta(l)]$ and 
$t=u[\delta(r)]$. 
When  we apply an arbitrary  equation $l\approx r\in E\cup E^{-1}$, we 
rename the variables of $l$ and $r$ such that  $l\in T_\S(X_{k+m})$ and  $r \in T_\S(X_k\cup X_{[k+m+1, k+m+\ell]})$ 
for some 
$k, m, \ell\geq 0$.

The  word problem  for a TES $E$ is the problem of deciding
for arbitrary $p, q\in \ts(X)$ whether $p\tthue E q$.
 The  ground word problem for $E$ is the 
 word problem restricted to ground terms $p$ and $q$.

For the notion  of a term rewriting system (TRS), see Section 4.2 in \cite{baanip}

{\bf  Knuth-Bendix   completion procedure.} 
We now briefly recall the basic Knuth-Bendix   completion procedure, see Section 7.1 in  \cite{baanip}.
The basic Knuth-Bendix   completion procedure starts with a TES $E$ and tries to find a convergent 
TRS $R$ that is equivalent to $E$. A reduction order $>$ is provided as an input for the procedure. 
Since the word problem is not decidable in general, a finite convergent TRS cannot always be obtained. 
In the basic Knuth-Bendix completion procedure this could be due to failure or to non-termination 
of completion.
In the initialization
phase, the basic completion procedure removes trivial identities of the form $s=s$ and tries to orient the remaining 
nontrivial identities. If this succeeds, then it computes all critical pairs of the TRS obtained. 
The terms in each critical pair
$\langle s, t \rangle$ are reduced to their normal forms $\hat{s}$ and $\hat{t}$. 
If the normal forms are identical, then this critical pair is joinable, and nothing needs to be done for it. 
Otherwise, the procedure tries to orient the terms $\hat{s}$ and $\hat{t}$ into the  rewrite rule 
 $\hat{s} \rightarrow \hat{t}$ with $\hat{s} > \hat{t}$
or $\hat{t} \rightarrow \hat{s}$ with $\hat{t} > \hat{s}$. 
In this way the procedure orients all instances of the terms $\hat{s}$ and $\hat{t}$ as well.
If this succeeds, then the new rule is added to the current rewrite system. This process is iterated until failure occurs or the rewrite system is not changed during a step of the iteration, that is, the system does not have non-joinable critical pairs. 

If the basic completion procedure applied to $(E, >)$ terminates succesfully with output  $R$, then $R$ is a finite convergent
TRS that is equivalent to $E$. In this case, $R$ yields a decision procedure for the word problem for $E$. 
If the basic completion procedure applied to $(E, >)$ does not terminate, then it outputs  an infinite 
 convergent TRS that is equivalent to $E$. In this case, the completion procedure can be used as a semidecision procedure 
 for the word problem for $E$. 

Assume that we want to decide for given terms $p, q\in \ts(X)$, whether $p\tthue E q$ holds.
We call the pair $(p,  q)$  the {\em goal}.
The  basic Knuth-Bendix completion procedure is independent of the goal. 
Hence, if  $p\tthue E q$ does not hold,  and the set $E$ of equations has no finite convergent system, then the basic  
Knuth-Bendix completion will run forever.
In the light of this observation,  
Lynch and Strogova \cite{lyn,ls} presented a goal-directed completion procedure based on SOUR graphs. 
Similarly to the basic Knuth-Bendix completion procedure, the  goal-directed completion procedure 
uses a  reduction order $>$. Unlike the  basic Knuth-Bendix completion procedure, it   uses some  inference rules. 
The main difference,  described in an intuitive  simplified way, is the following.  Along  the completion procedure, we try to construct  a rewrite system $R$ and a conversion 
\begin{equation}\label{utcabal}
p = r_1\thue R r_2 \thue R \cdots \thue R r_n = q, \; n\geq 1
\end{equation}
in a nondeterministic way.  We compute and orient critical pairs and 
control  the completion process keeping in our mind that 
the  rules of $R$ 
should be applicable  along a conversion (\ref{utcabal}). When orienting the equations into rules along the completion process, 
we do not put a rule in $R$ if it is not 
 applicable  along a conversion (\ref{utcabal}).  If we do not find a conversion (\ref{utcabal}), the goal-directed completion procedure 
detects that $(p, q)\not \in \tthue E$,
outputs 'no' and halts.
 Consider the following example.
 Let ranked alphabet $\S$ consist of the unary symbols $f$, $g$
 and the nullary symbols
$\$$, $\#$. 
Consider the variable preserving TES $E=\{\, ffx\approx gfx\, \}$. We raise the problem whether
$\$\tthue E \#$. The basic Knuth-Bendix completion procedure runs forever on this example  \cite{lyn}. 
Along the goal oriented completion procedure,
we find  no  rewrite   rule such that it is  applicable along a  conversion 
$\$ = r_1\thue R r_2 \thue R \cdots \thue R r_n = \#$, $n\geq 1$.
Therefore, the goal-directed completion procedure detects that $(\$, \#)\not \in \tthue E$,
outputs 'no', and halts  \cite{lyn}.

We now adopt  a  more detailed description of the  goal-directed completion procedure.  \cite{ls}
The  goal-directed completion procedure 
uses a  reduction order $>$ and computes critical pairs equipped with equational and ordering constraints, and 
constructs  a graph.  
``The goal-directed completion procedure has  two phases. The first phase is the compilation phase.
 In this phase, all the edges and the recursive constraints labelling each edge are created. This phase also takes into account the goal to be solved.
Importantly, this phase takes only polynomial time, because there are only polynomially many edges in the graph. The result of this phase
is a constrained tree automaton representing a schematized  version of the completed system, and a set of constraints representing potential solutions to the goal. The constraints that are generated are the equational constraints representing the unification problems, and ordering constraints arising from the critical pair inferences. 
 
The second phase is the goal solving (or constraint solving) phase. In this phase, the potential solutions to the goal are solved in order to determine whether they are actual solutions of the goal. 
This phase can take infinitely long, since the constraints are recursive. 
 Step by step a constraint is rolled back, based on which edges it is created from, and the equational and ordering constraints are  solved along the way. 
 In some cases, the ordering constraints cause the recursion to halt, and therefore the constraints are completely solved.
 The procedure is truly goal oriented, because only a polynomial amount of time is spent compiling the set of equations. The rest of the time is spent  working backwards from the goal to solve the constraints. If the procedure is examined more closely, we see that the second phase of the procedure is exactly a backwards process of completion.
 A schematization of an equation in the completed system is applied to the goal, step by step until it rewrites to an
 identity. At the same time, the schematized equation that is selected is worked backwards until we reach the original 
 equations from which it is formed.'' \cite{ls}

See Section 7.2 in \cite{baanip}
for an  improved version of the  
Knuth-Bendix completion procedure described by a set of inference rules. 
A detailed description of the unfailing  Knuth-Bendix completion procedure can be found in \cite{bdp}. 

{\bf Ground term equation systems and rewriting systems.} 
A ground term equation system (GTES) $E$  over a ranked alphabet $\S$
is a finite binary relation on $\ts$.  Elements $(l,r)$ of $E$ are 
called equations and are denoted
by $l\approx r$. 
The reduction relation $\red E \subseteq \ts(X) \times \ts(X)$ is defined as 
follows. 
For any  ground terms $s, t\in \ts$, $s \red E t$ if there is a pair $l\approx r$
in $E$ and 
a context $u\in C_\S(X_1)$ such that $s=u[l]$ and 
$t=u[r]$. It is well known that the relation $\tthue E$ is a 
congruence on the term algebra ${\bf TA}$ \cite{vag2}. We call 
$\tthue E$ the congruence induced by $E$.
The size of $E$ is defined as the number of occurrences of symbols in the set. 
$sub(E)=\{\, sub(l) \mid l\approx r\in E\cup E^{-1} \, \}$.  
Clearly,  $\tthue E \cap (sub(E)\times sub(E))$ is an equivalence relation on $sub(E)$. 
The  word problem  for a GTES $E$ is the problem of deciding
for arbitrary $p, q\in \ts$ whether $p\tthue E q$.

A ground term
rewrite system (GTRS) over a ranked alphabet $\S$ is 
a finite subset $R$ of $\ts\times \ts$. The elements of $R$
are called rules and a rule $(l,r)\in R$ is written in
the form $l\rightarrow r$ as well. Moreover, we
say that $l$ is the left-hand side and $r$ is the right-hand
side of the rule $l\rightarrow r$. 
$lhs(R)=\{l\mid l\rightarrow r \in R\, \}$,
 $rhs(R)=\{r\mid l\rightarrow r \in R\, \}$.
$sub(R)=\{\, sub(l) \mid l\in lhs(R) \, \}\cup
\{\, sub(r) \mid r\in lhs(R) \, \}$.  

The reduction relation $\red R \subseteq \ts(X) \times \ts(X)$ is defined as 
follows. 
For any  ground terms $s, t\in \ts$, $s \red R t$ if there is a pair $l\approx r$
in $E$ and 
a context $u\in C_\S(X_1)$ such that $s=u[l]$ and 
$t=u[r]$. 
 Here we say that
$R$ rewrites $s$ to $t$ applying the rule $l\rightarrow r$.
A GTRS $R$ is {\em equivalent to} a GTRS $E$, if 
$\tthue R=\tthue E$ holds.

$\textit{IRR}(R)$ denotes the set of all ground terms irreducible by  $R$.
A GTRS $R$ is reduced if for every rule $u\rightarrow v$ in $R$,
$u$ is  irreducible with respect to $R-\{\, u\rightarrow v\,\}$
and $v$ is irreducible with respect to $R$. For a reduced GTRS $R$, 
$\textit{IRR}(R)\cap sub(R)=sub(R)-lhs(R)$,
and   
$sub(R)-lhs(R)$ is a set of
representatives for $sub(R)/ \tthue R$, see  Theorem 3.14 on page 162 in  \cite{vag1}.

We say that a  GTRS $R$ is  confluent, locally confluent, terminating, or 
convergent, if $\red R$ has the corresponding property.

We recall the following important result.
\begin{prop}\rm\label{mozart} \cite{sny} Any reduced GTRS $R$ is
convergent.
\end{prop}
\begin{prop}\rm\label{bartok}
For a reduced GTRS $R$, one can reduce a ground  term $t\in \ts$ to its normal form in linear time of $size(t)$.
We traverse the  term $t$ in postorder. When visiting a position $\alpha$, we reduce the 
subterm $t/\alpha$ of $t$ at $\alpha$  to is normal form
$t/\alpha\hspace{-1mm}\downarrow_{R}$. 
\end{prop}

We say that a GTRS $R$  is equivalent to a GTES $E$ if $\tthue R=\tthue E$. 
\begin{prop} \label{kairo} {\rm \cite {sny}}
For a GTES $E$
one can effectively construct  an equivalent  reduced GTRS $R$ in $O(n \, log \, n)$ time.
Here $n$ is the size of $E$.
\end{prop}
\pr We briefly recall Snyder's \cite{sny} fast ground completion algorithm.
We run a congruence closure algorithm for $E$ over the subterm graph of $E$
\cite{dst,nelopp}. In this way we
 get  the representation of the equivalence relation  $\tthue E\cap (sub(E)\times sub(E))$.
We  compute a set $REP$ of representatives for  $sub(E)/\tthue E $. Then
we construct a reduced GTRS $R$ over $\S$ as follows. We put 
the rewrite rule $l\rightarrow r$  in $R$ if

$\bullet$ $l=f(p_1, \ldots, p_m)$ for some $f\in \S_m$,  $m\geq 0$,
and $p_1,\ldots,p_m\in REP$,

$\bullet$ $r\in REP$, 

$\bullet$  $l\neq r$ and $l \tthue E r$.

\hfill $\b$

We can decide the word problem of a GTES $E$ applying a congruence closure algorithm 
\cite{dst,nelopp} for the GTES $E_1=E\cup \{\, p\approx p, q\approx q \, \}$ and then 
examine whether $p, q$ are in the same class of the 
 equivalence relation  $\tthue {E_1}\cap (sub(E_1)\times sub(E_1))$.
 Assume that we want to solve the word problem of a fixed  GTES $E$ for varying terms  $p, q$. Then we compute a 
convergent GTRS over $\S$ equivalent to $E$ \cite{gnprs,ls,plasat,sny}. 
We compute $p \hspace{-1mm}\downarrow_{R}$
and $q \hspace{-1mm}\downarrow_{R}$, and compare them.  If 
$p \hspace{-1mm}\downarrow_{R}= q \hspace{-1mm}\downarrow_{R}$, then  $p\tthue E q$. Otherwise, $(p, q) \not \in \tthue E$. 
By Proposition \ref{bartok}, we can decide the word problem of $E$ in linear time.
 We can also extend the signature. We introduce constants for the  equivalence  classes 
of  $\tthue E\cap (sub(E)\times sub(E))$. Then  
we can construct  in $O(n \, log \, n)$ time a reduced GTRS over the extended signature such that
$p \hspace{-1mm}\downarrow_{R}= q \hspace{-1mm}\downarrow_{R}$
 if and only if  $p\tthue E q$. 
By Proposition \ref{bartok}, we can decide the word problem of $E$ in linear time.
  Finally, assume that we want to solve the word problem of a fixed  GTES $E$ for a fixed term $p$ and varying term 
  $q$.
Then we can construct  in $O(n \, log \, n)$ time a deterministic tree automaton 
recognizing the $\tthue E$-class of $p$
\cite{vag1}.

\noindent
For other completion algorithms on GTRSs see \cite{fulvag3,plasat}. 
For further results on GTRSs  see  \cite{vag2}.
Proposition \ref{mozart} and Proposition \ref{kairo} imply the following well known result. 
\begin{prop} \label{memphis} {\rm \cite {sny}}
For a GTES $E$  and ground terms $p$, $q$, 
one can decide whether $p\tthue E q$. 
\end{prop}

\section{Reachability starting from a term attached to a context}\label{rur}
Let $R$ be a  reduced GTRS over $\S$ and 
$s, t\in \textit{IRR}(R)$. We say that
 $R$  reaches $t$ starting from $s$ attached to some context,  if  there is a $u\in C_\S$ 
such that $u[s]\tred R t$. Let $\textit{RAC}(s)$ denote the set of all terms $t\in \textit{IRR}(R)$ which are reachable by 
$R$  starting from $s$ attached to some context. 
\begin{exa}\label{sportos}{\em 

Let $\S=\S_0\cup \S_1$, $\S_0=\{\, 0, 1\, \}$, and $\S_2=\{\, f\, \}$.
Let GTRS $R$ consist of the equations
$f(0, 0)\rightarrow  0$ and 
$f(0, 1)\rightarrow  1$.
Clearly $R$ is reduced.
Then each element of $\textit{IRR}(R)$  containing $0$ is in $ \textit{RAC}(0)$.
 For example, $f(f(1, 0), 1)\in \textit{RAC}(0)$, because $f(f(1, \diamond), 1)\in C_\S$ and 

 $f(f( 1, \diamond), 1)[0]= f(f(1, 0), 1)\tred R f(f(1, 0), 1)$. 

\noindent Furthermore, $1 \in  \textit{RAC}(0)$, because 

$f(\diamond, 1)[0]= f(0, 1) \red R  1$. 

Thus each element of $\textit{IRR}(R)$  containing $1$ is in $ \textit{RAC}(0)$.
Consequently, $\textit{IRR}(R) = \textit{RAC}(0)$.

\hfill $\b$ } 
\end{exa}

\begin{lem} \label{szamol} Let $R$ be a  reduced GTRS over $\S$. For any 
 $s\in sub(R)-lhs(R)$, we can effectively compute  $\textit{RAC}(s)\cap (sub(R)-lhs(R))$.
\end{lem}\pr 
Let $\textit{RAC}_0=\{\, s\, \}$.
For each $i\geq 0$, let $\textit{RAC}_{i+1}$ consists of all elements $t$, where

$\bullet$ $t\in \textit{RAC}_i$ or

$\bullet$ $t\in sub(R)-lhs(R)$ and   there is a rule $ f(t_1, \ldots, t_m)\rightarrow t$
 in $ R$ for some  $f\in \S_m$,  $t_1, \ldots, t_m\in sub(R)-lhs(R)$, such that   $t_j \in \textit{RAC}_i$
for some $1\leq j \leq m$,  or 

$\bullet$ $t\in sub(R)-lhs(R)$ and 
$t=f(t_1, \ldots, t_m)$ for some $ f\in \S_m$,   $t_1, \ldots, t_m\in sub(R)-lhs(R)$,   and $ t_j \in \textit{RAC}_i $
for some $1\leq j \leq m$. 
Then 
\begin{equation}\label{rakoczi}
\textit{RAC}_i \subseteq \textit{RAC}_{i+1} \subseteq \textit{RAC}(s) \cap (sub(R)-lhs(R)) \mbox{ for } i\geq 0\, . 
\end{equation}
Hence there is an integer $0\leq \ell \leq card(sub(R)-lhs(R))$ such that 
$\textit{RAC}_\ell= \textit{RAC}_{\ell+1}$. Then 
\begin{equation}\label{tokoli}
\textit{RAC}_\ell= \textit{RAC}_{\ell+k} \mbox{  for } k\geq 1\, . 
\end{equation}
Hence 
\begin{equation}\label{kapu}
\textit{RAC}_\ell\subseteq  \textit{RAC}(s)\cap  ( sub(R)-lhs(R))\, .
\end{equation}
To show the reverse inclusion, we need the following. 
\begin{cla}\label{zuza}For any  $u\in C_\S$ of height $n\geq 0$ and $t\in  sub(R)-lhs(R)$,  if $u(s)\tred R t$, then 
$t\in \textit{RAC}_n$. \end{cla}
\pr By induction on $n$.

\hfill $\b$

\noindent By (\ref{rakoczi}), (\ref{tokoli}), and Claim \ref{zuza}, $\textit{RAC}(s)\cap   ( sub(R)-lhs(R))\subseteq  \textit{RAC}_\ell$. 
By (\ref{kapu}),  $$ \textit{RAC}(s)\cap   ( sub(R)-lhs(R))= \textit{RAC}_\ell\, .$$
We compute the sets 
$\textit{RAC}_0, \textit{RAC}_1, \ldots, R_{ card(sub(R)-lhs(R)) }$.  In this way we obtain the integer $\ell$ and $\textit{RAC}(s)\cap   ( sub(R)-lhs(R))$.

\hfill $\b$

\begin{lem} \label{alakit}For any reduced GTRS $R$ and 
 $s, t\in \textit{IRR}(R)$, 
 $R$  reaches $t$ starting from $s$ attached to some context
if and only if 
 
(i) $t= u[s]$ for some 
$u\in C_\S$ or  

(ii) $s\in (sub(R)-lhs(R))$, and  there are  $u\in C_\S$ and  $r \in rhs(R)$ such that  $t= u[r]$ 
and  $R$  reaches $r$ starting from $s$ attached to some context.

\end{lem}\pr 
$(\Rightarrow)$ Assume that 
 $R$  reaches $t$ starting from $s$ attached to some context.
Then   there is $u\in C_\S$ 
such that $u[s]\tred R t$.
If $u[s]= t$,  then   (i) holds.
 Otherwise, $u[s] \rightarrow^+ _R t$. Hence there are   
$v_1, v_2, z\in C_\S$ and a rule $l\rightarrow r$ in $R$  such that 
$u[s]=v_1[z[s]]\tred R v_1[l]   \red R v_1[r]\tred R v_2[r]=t$, where

(a)  $u= v_1[z]$, 

(b)   $z[s]\tred R l$, 

(c)  $l \rightarrow r\in R$, 

(d)   $v_1\tred R v_2$  over the ranked alphabet $\S \cup \diamond$. 

\noindent 
Hence $t=v_2[r]$, $v_2\in C_\S$,  $r\in rhs(R)$. 
By (b), $s\in sub(l)$ or $s\in sub(l_1) $ for some $l_1\in LHS(R)$. Recall that $s\in \textit{IRR}(R)$. 
Hence $s\in (sub(R)-lhs(R))$.


$(\Leftarrow)$ If (i) holds, then  $R$  reaches $t$ starting from $s$ attached to some context.

Assume that (ii) holds.  Then  there is $z\in C_\S$
such that $z[s]\tred R r$. Consequently $(u[z])[s]=u[z[s]]\tred R u[r]=t$. 
Hence 
 $R$  reaches $t$ starting from $s$ attached to some context.

\hfill $\b$

Lemma \ref{szamol} and Lemma \ref{alakit} imply the following result. 
\begin{prop} \label{eldont} For any 
 $s, t\in \textit{IRR}(R)$, we can decide whether  $R$  reaches $t$ starting from $s$ attached to some context.
\end{prop}

\section{Two trivial semi-decision procedures}\label{led} 
We present the well known trivial  semi-decision procedure $\textit{PRO1}$
 for the ground word problem of  variable preserving TESs. We give examples when $\textit{PRO1}$ is efficient. Then we present 
the trivial  semi-decision procedure $\textit{PRO2}$  for the ground word problem of TESs. Note that $\textit{PRO2}$ is a straightforward
generalization of $\textit{PRO1}$.

\vspace{ 3mm} 

\noindent
{\bf Procedure} $\textit{PRO1}$
\noindent Input:  A variable preserving  TES $E$ over the ranked alphabet $\S$ and  ground terms $p, q\in \ts$.
 
\noindent  Output:  'yes' if $p \tthue E q$, 
            'no'  or undefined otherwise.

\noindent   Let $U_0 = \{\, p \,\}$, $V_0 = \{\, q \,\}$, $i = 0$. 
  
\noindent  
repeat

  $i := i+1$; 
      
 $U_i := U_{i-1} \cup \{ \, s | \mbox{ there is } u\in U_{i-1} \mbox{ such that } 
u \thue E  s \, \}$;

 $V_i := V_{i-1} \cup \{ \, s | \mbox{ there is } u\in V_{i-1} \mbox{ such that } 
 u \thue E  s\,\}$;

   \noindent
until ($U_i = U_{i-1}$ or   $V_i = V_{i-1}$) or 
$U_i \cap V_i$ is not empty;

if $U_i \cap V_i$ is not empty
            
then  begin output 'yes';  halt end;

          \noindent   output 'no'; 

          \noindent     halt

\vspace{3mm} 

\noindent
For any  variable preserving  TES $E$ and  ground term $u$, the set 
$\{\, s| u \thue E s\,  \}$
 is finite 
and then effectively computable. Thus for every  $i\geq 0$, $U_i$ and $V_i$,  are finite and  can be computed effectively.
Hence the above procedure can be implemented. Clearly, $\textit{PRO1}$ outputs 'yes' and halts 
if and only if $p \tthue E q$. If  $\textit{PRO1}$ outputs 'no' and halts, then $(p, q)\not\in \tthue E$.

We adopt the following example of Lynch \cite{lyn}. 
\begin{exa}\label{futo} {\em  Let  $\S=\S_0\cup \S_1$, $\S_0=\{\,\$, \#\, \}$, $\S_1=\{\,f, g\, \}$. 
Consider the TES $E=\{\, ffx\approx gfx\, \}$. We raise the problem whether
$\$\tthue E \#$. On the one hand,  the basic  Knuth-Bendix completion procedure runs forever on this example  \cite{lyn}. 
On the other hand, the goal-directed completion procedure outputs
'no' and halts \cite{lyn}. 
It is still open whether  the goal-directed completion procedure halts on the TES $E$ and any goal  \cite{lyn}.

 Observe that for each $u\in \ts$, the set $\{\, s\mid u\tthue E s\, \}$ is finite. Hence for any $p, q\in \ts$, $\textit{PRO1}$ outputs the correct answer and  halts. 
For this example, $\textit{PRO1}$  is more efficient than
the   basic Knuth-Bendix completion procedure, and is at least as efficient 
as  the goal-directed completion procedure \cite{lyn, ls}.

\hfill $\b$ }
\end{exa}

\begin{exa}\label{kiraly} {\em  Let  $\S=\S_0\cup \S_2$, $\S_0=\{\,\star, \$, \#\, \}$, and $\S_2=\{\,f\, \}$.
We define the terms $comb_i\in \ts(X_i)$,  $i\geq 1$,  as follows. 
Let $comb_1=f(x_1,\star)$, $comb_{i+1}=f(x_1, comb_i[ x_2, \ldots, x_{i+1}])$ for $i\geq 1$.
For example, $comb_3 = f(x_1, f(x_2, f(x_3, \star)))$. 
Let $n\geq 1$, $p=comb_{2n}[\#,  \ldots, \#]$, and $q=comb_{2n}[\$,  \ldots, \$]$.
We run  procedure $\textit{PRO1}$ on  the  TES $E=\{\, \#\approx \$\, \}$ and the ground terms  $p$ and $q$. 
Then  

$card(U_{i})=card(V_{i}) =
\left(\begin{array}{c}
2n\\
i\\
\end{array}\right)+
\left(\begin{array}{c}
2n\\
i-1\\
\end{array}\right)+ \cdots +\left(\begin{array}{c}
2n\\
1\\
\end{array}\right)
$  for $i=1, \ldots n$, 

$U_{i} \cap V_{i}= \emptyset$ for $i=0, 1, \ldots n-1$, and 

$comb_{2n}[\#, \ldots,\#, \$,  \ldots,  \$]\in U_{n} \cap V_{n}$.

\noindent
Hence in the $n$th    step, $ \textit{PRO1}$ outputs 'yes' and halts. 

\hfill $\b$  }
\end{exa}

\begin{exa}\label{netes}{\em 
We present Ceitin's \cite{cei,ms} semi-Thue system  as a TES. Let  $\S=\S_0\cup \S_1$, $\S_0=\{\,\$\, \}$, and $\S_1=\{\,a, b, c, d, e \, \}$. $E$ consists of the equations 

$acx_1 \approx cax_1$, $adx_1 \approx dax_1$, $bcx_1 \approx cbx_1$,
 $bdx_1 \approx dbx_1$, 

$ecax_1\approx ce x_1$, $edbx_1 \approx dex_1$,

$cdcax_1 \approx cdcaex_1$, $caaax_1 \approx aaax_1$,  $daaax_1 \approx aaax_1$.

\begin{prop}\label{halasz}{\em \cite{cei,ms}} It is undecidable 
 for an arbitrary given ground term $t \in \ts$ whether  $t \tthue E a^3\$$. 
\end{prop}

We run    procedure $\textit{PRO1}$
on   the  TES $E$ and the ground terms
 $p=a^3\$$ and  $q=edb\$$.  We    compute as follows.
$U_0=\{\, p\, \}$, $V_0=\{\, q\, \}$, 
$U_1=\{\, a^3\$,ca^3\$, da^3\$ \, \}$, $V_1=\{\, edb\$,ebd\$, de\$ \, \}$, $U_2=\{\, a^3\$,ca^3\$, da^3\$, cca^3\$,   cda^3\$,dca^3\$, dda^3\$, acaa\$, adaa\$  \, \}$, $V_2=V_1$. 
Now procedure $\textit{PRO1}$  outputs 'no' and  halts. 

Let 
$n\geq 1$, $p=(bd)^{2n}\$$, and $q=(db)^{2n}\$$. 
We apply  procedure $\textit{PRO1}$ to   TES $E$ and ground terms  
$p$ and $q$.   We    compute as follows.

\noindent
$U_0=\{\, p\, \}$, $V_0=\{\,q\, \}$,

\noindent
$U_1=\{\,p, db(bd)^{2n-1}\$,\ldots, (bd)^{2n-1}db  \$  \, \}$,

\noindent
$V_1=\{\, q, bd(db)^{2n-1}\$,\ldots, (db)^{2n-1}bd\$,
 \, \}$,

\noindent
$U_2=U_1\cup \{\  dbdb(bd)^{2n-2}\$, dbbddb(bd)^{2n-3}\$,
\ldots,  (bd)^{2n-2} dbdb\$ \, \}$,

\noindent
$V_2=V_1\cup \{\, bdbd(db)^{2n-2}\$, bd db bd (db)^{2n-3}\$,\ldots,
(db)^{2n-2}bdbd\$ \, \}$,

$\ldots$ .

\noindent 
Observe that 
$U_{i}\cap V_{i}=\emptyset$ for $i=0, 1, \ldots, n-1$. 
Clearly, 
$(bd)^{n}(db)^{n}\$\in U_{n}\cap V_{n}$. After computing  
$U_{n}$ and $V_{n}$, 
  procedure $\textit{PRO1}$ outputs 'yes' and halts. 

\hfill $\b$
}
\end{exa}

\begin{exa}\label{cipo}{\em  We continue Example \ref{netes}.
Let $p\in \ts$ be arbitrary such that  symbols $a$ or $c$ appear in $p$.
  Let $q\in \ts$ such that $a, c$ do not appear in $q$.
That is, only the constant $\$$ and the symbols $b$, $d$, or  $e$ appear in $q$.

Observe that the left-hand side and the right-hand side of the  fourth and sixth rules do not contain  $a$ or $c$.
Both sides of all  other rules contain  $a$ or $c$. Hence 
for  any reduction sequence  
 
\noindent
$p \red R p_1 \red R p_2\red \cdots \red R p_n$, $n\geq 1$,  for any $1\leq i \leq n$, the term $p_i$ contains the constant 
$\$$ and at least one $a$ or $c$. 
Furthermore, along  any reduction sequence  
 $q \red R q_1 \red R q_2\red \cdots \red R q_n$, $n\geq 1$,   
 we only use the fourth and sixth equations.
Consequently,  the set $\{\, v \in \ts \mid q \tthue E  v\, \}$ is finite. 
Furthermore  neither $a$ nor  $ c$ appears in any element  of  the set $\{\, v \in \ts \mid q \tthue E  v\, \}$. 
Thus 
\begin{equation}\label{tabori}
(p, q) \not \in \tthue E\,,
\end{equation}
and  $U_i \cap V_i=\emptyset$ for $i\geq 0$. Thus procedure $\textit{PRO1}$ outputs 'no' and halts
on the input $E$, $p$, $q$. 

\hfill $\b$
}
\end{exa}

\begin{exa}\label{tavasz} {\em 
Let  $\S=\S_0\cup \S_1$, $\S_0=\{\,a\, \}$, and $\S_1=\{\, f
\, \}$. 
TES $E$ consists of the equation $ffx \approx x$. 
We run   procedure $\textit{PRO1}$ on   TES $E$ and ground terms  $p=a$
and $q=fa$.  We    compute as follows.

\noindent
$U_0=\{\, a\, \}$, $V_0=\{\,fa \, \}$, 

\noindent
$U_1=\{\,a, ffa \, \}$, $V_1=\{\, fa, f^3a\, \}$,

\noindent
$U_2=\{\,a, ffa, f^4a \, \}$, $V_2=\{\, fa, f^3a, f^5a  \, \}$,$\ldots$.

\noindent
$U_0\subset U_1\subset U_2 \subset \cdots$,

\noindent
$V_0\subset V_1\subset V_2 \subset \cdots$, and

\noindent           
$U_i \cap V_i=\emptyset$ for $i\geq 0$.

\noindent
Hence procedure $\textit{PRO1}$ does not halt. 

\hfill $\b$ }
\end{exa}


To present the   
semi-decision procedure   $\textit{PRO2}$,  
we define the sets $U_i\subseteq  \ts$,  $i\geq 0$,  by recursion. 
  Let $U_0 = \{\, p \,\}$.  Let  $i\geq  1$.    
We put all elements of $U_{i-1}$ in $U_i$. Moreover, we put in $U_i$ all
$s\in \ts$ such that   

$\bullet$  $l'\approx r'$ is a 
  ground instance 
 of some  equation  $l\approx r$  in  $E\cup E^{-1}$ obtained by 
 substituting arbitrary ground terms of 
height less than or equal to $ i-1$
for all variables that do not appear in $l$,

$\bullet$ 
 $v\in C_\Sigma$,  

$\bullet$ 
$ v[l']\in U_{i-1}$ and   $s=v[r']$.

\noindent 
We define $ V_i \subseteq  \ts$,   $i\geq 0$, symmetrically to  $U_i$,  $i\geq 0$. 
Clearly for every  $i\geq 0$, $U_i$ and $V_i$  are finite and
can be computed effectively. Note that there may be an $i\geq 1$ such that  $U_i=U_{i+1}$ and 
$ U_{i+1}\subset U_{i+2}$. 

\begin{exa}\label{makaroni}{\em 
Let $\S=\S_0\cup \S_1$, $\S_0=\{\, 0, 1\, \}$, and $\S_2=\{\, f\, \}$.
Let TES $E$ consist of the  equations

$f(x_1, x_1)\approx 0$, 
$f(0, x_1)\approx  x_1$.

\noindent Let $p=f(1, 0)$ and $q=f(1, f(1, 1))$.
Then 

\noindent
$U_0=\{\, f(1, 0)\, \}$, $V_0= \{\, f(1, f(1, 1))\, \}$,

\noindent
$U_1=\{\,f(1, 0),  f(f(0, 1), 0), f(1, f(0, 0)),  f(1, f(1, 1))    \, \}$,

\noindent
$V_1= \{\, f(1, f(1, 1)),  f(f(0, 1) , f(1, 1)),  f(1, 0),
f(1, f(f(0, 1) , 1)), 
 f(1, f(1, f(0, 1) ))
 \, \}$.

\hfill $\b$
}
\end{exa}

\vspace{ 3mm} 

\noindent
 {\bf Procedure} $\textit{PRO2}$
Input:  A TES $E$ over the ranked alphabet $\S$ and  ground terms $p, q\in \ts$.

\noindent  Output:  'yes' if $p \tthue E q$, 
             undefined otherwise.

\noindent  1 \hspace{1.7mm}  $i := i+1$;
   
                        compute $U_i$ and $V_i$;

if  $U_i \cap V_i$ is not empty then begin 
             output 'yes'; halt end; 

goto 1 

\vspace{ 3mm} 

\noindent 
  $\textit{PRO2}$ outputs 'yes' and halts 
if and only if $p \tthue E q$.

\begin{exa}\label{tejfel}{\em 
We continue  Example \ref{makaroni}. 
We run   procedure $\textit{PRO2}$ on   TES $E$ and ground terms  $p, q$. 
 We    compute as follows. We compute $U_0$ and $V_0$. We observe that 
 $U_0 \cap V_0$ is empty. Then we compute $U_1$ and $V_1$. We observe that 
 $U_1 \cap V_1$ is not empty. Procedure $\textit{PRO2}$ outputs 'yes' and halts. 

\hfill $\b$
}
\end{exa}

\section{Semi-decision procedure for the ground
             word problem of variable preserving TESs }\label{fo}
We present the semi-decision procedure  $\textit{PRO3}$ for the ground
word problem of variable preserving TESs,  and show its correctness.  
$\textit{PRO3}$ is an improvement of $\textit{PRO1}$. The starting idea is the following. For each $i\geq 1$, 
we construct the GTES $P_i$ using  those instances of equations in $E\cup E^{-1}$ which are applied to compute the set $U_i$. We
 improve this construction by defining $P_{i}$, $i\geq 2$,  as the set of all  instances of equations in $E\cup E^{-1}$ which 
can be applied to elements of $\{\, s\in \ts \mid p \tthue {P_{i-1}} s\, \}$  rather than to the elements of $U_{i-1}$. Furthermore,  
we define the GTES $Q_i$ symmetrically. 
We give examples when procedure $\textit{PRO3}$ is more efficient  than procedure $\textit{PRO1}$.

Let $E$ be a  variable preserving TES over $\S$, and let $p, q\in \ts$. 
We define the GTESs $P_i$ and the reduced GTRSs $R_i$, $i \geq 1$, 
 over $\S$ as follows.

For each equation  $l\approx r$ of $E\cup E^{-1}$ 
with $l, r \in\ts(X_m)$, $m\geq 0$, and for any  $u\in C_\S$, 
$u_1, \ldots, u_m\in\ts$, 
if  
$p=u[l[u_1, \ldots, u_m]]$
then we put the  equation
$l[u_1, \ldots, u_m] \approx r[u_1, \ldots, u_m]$
 in $P_1$.
Applying Snyder's algorithm   we compute a reduced GTRS $R_1$ equivalent to the GTES 
$P_1$, see Proposition \ref{kairo}.

Let $i\geq 1$. 
(a)  We put 
each element  of $R_i$ into $P_{i+1}$. 

(b)
For each equation  $l\approx r$ of $E\cup E^{-1}$, $l, r \in\ts(X_m)$, $m\geq 0$, 
for any $u_1, \ldots,  u_m\in (sub(R_i)-lhs(R_i))\cup sub( p\hspace{-1mm}\downarrow_{R_i})$, if 
$R_i$  reaches $p\hspace{-1mm}\downarrow_{R_i}$ starting from $l[u_1,  \ldots, u_m]\hspace{-1mm}\downarrow_{R_i}$ attached to some context, and
$l[u_1, \ldots, u_m]\hspace{-1mm}\downarrow_{R_i} \neq  r[u_1, \ldots, u_m]\hspace{-1mm}\downarrow_{R_i} $, 
then we put the equation  
$l[u_1, \ldots, u_m] \approx r[u_1, \ldots, u_m]$
in $P_{i+1}$.

If $P_{i+1}=R_i$, then let $R_{i+1}=R_i$. Otherwise, 
applying Snyder's algorithm,   we compute a reduced GTRS $R_{i+1}$ equivalent to the GTES $P_{i+1}$.

When misunderstanding may arise, we denote $R_i$ as  $R_{P_i}$.
We define the GTESs $Q_i$, $i\geq 1$, symmetrically to the GTESs $P_i$, $i\geq 1$.
Applying Snyder's algorithm,   we compute a reduced GTRS $R_{P_i \cup Q_i}$ equivalent to the GTRS
$R_{P_i} \cup R_{Q_i}$ for $i\geq 1$. 

We illustrate our concepts and results by the following  example. 
\begin{exa}\label{takacs}{\em 

Let $\S=\S_0\cup \S_1\cup \S_2$, $\S_0=\{\, \$, \#\, \}$, $\S_1=\{\, e, f, g, h\, \}$, and $\S_2=\{\, d\,  \}$. 
Let the TES $E$ consist of the equations

$\#\approx \$$,  \, 
$g\$\approx h\$$,   \, 
$d(hx_1, hx_1)\approx hx_1$,   \, 
$efhx_1\approx hx_1$. 

\noindent 
Observe that $E$ is  variable preserving. 
\noindent
Let $p=efg\#$, $q=d(h\#,h\# )$. 

\vspace{3 mm} 


First we compute the GTES $P_i$, $i\geq 1$.
\noindent GTES $P_1$  consists of the equation  $\#\approx \$$.
Let $\Theta$ stand for $\tthue {P_1}\cap (sub(P_1) \times sub(P_1))$.
Then 
$sub(P_1)/\Theta=\{\, \{\, \#, \$\, \}\, \}$ and $\{\, \$ \, \}$ 
is a set of representatives for $sub(P_1)/\tthue {P_1}$.
GTRS $R_1$ consists  of the rule
$\#\rightarrow \$$.

GTES $P_2$ 
consists of the equations
$\#\approx  \$$, \hspace{ 3mm} 
$g\$\approx h\$$. 
Let $\Theta$ stand for $\tthue {P_2}\cap (sub(P_2) \times sub(P_2))$.
Then 
$sub(P_2)/\Theta=\{\, \{\, \#, \$\, \}, \{\, g\#, g\$, h\#, h\$    \, \}     \, \}$ and 
$\{\, \$ , h\$\, \}$ 
is a set of representatives for $sub(P_2)/\tthue {P_2}$.
GTRS $R_2$ consists  of the rules
$\#\rightarrow \$$, 
$g\$\rightarrow  h\$$.


\noindent
GTES $P_3$ 
consists of the equations

$\#\approx  \$$, \, 
$g\$\approx h\$$,  \, 
$h\$ \approx d(h\$, h\$)$,  \, 
$h\$ \approx   efh\$ $.

\noindent
Let  $\Theta$ stand for $\tthue {P_3}\cap (sub(P_3) \times sub(P_3))$.
Then 

 $sub(P_3)/\Theta=\{\, \{\, \#, \$\, \},  \{\, g\#, g\$, h\#, h\$,    d(h\$, h\$),  efh\$  \},  \{\,  fh\$\, \}\, \}$

\noindent and 
$\{\, \$, h\$, fh\$ \, \}$ 
is a  set of  representatives for $sub(P_3)/\tthue {P_3}$.
$R_3$ consists  of the rules

$\# \rightarrow \$$,  \, 
$g\$\rightarrow h\$$,  \, 
$ d(h\$, h\$)\rightarrow h\$ $,  \, 
$efh\$ \rightarrow h\$$.

\vspace{3 mm}

\noindent $P_4=R_3$ and $R_4=R_3$. Furthermore,   $P_i=R_3$ and   $R_i=R_3$ for $i\geq 4$.

Second, we compute the  GTESs $Q_i$, $i\geq 1$.  
 GTES $Q_1$  consists of the  equations
$\# \approx  \$$, 
$  d(h\#, h\# )  \approx  h\#  $.
GTRS $R_{Q_1}$ consists  of the rules
$\#\rightarrow \$$, 
$d(h\$, h\$)\rightarrow  h\$$.

 GTES $Q_2$  consists of the  equations
$\# \approx  \$$, 
$  d(h\$, h\$ ) \approx   h\$ $,
$efh\$\approx h\$$.

GTRS $R_{Q_2}$ consists  of the rules
$\#\rightarrow \$$, 
$ d(h\$, h\$ ) \rightarrow    h\$ $,
$efh\$ \rightarrow  h\$$.

 Observe that  $R_{Q_2}=Q_i= R_{Q_i} $ for $i\geq 3$.

\noindent  $R_{P_1 \cup Q_1}=R_{P_1}$, $R_{P_2 \cup Q_2}=R_{P_2}\cup R_{Q_2} $, and 
$R_{P_3 \cup Q_3}= R_{P_3}$. Then 

$p\hspace{-1mm}\downarrow_{R_{P_1 \cup Q_1}} =efg\$$, 
 $q\hspace{-1mm}\downarrow_{R_{P_1 \cup Q_1}}=h\$$, 

$p\hspace{-1mm}\downarrow_{R_{P_2 \cup Q_2}} =h\$$, 
$q\hspace{-1mm}\downarrow_{R_{P_2 \cup Q_2}} =h\$$. 

\hfill $\b$ }
\end{exa}

We get the following result
by direct inspection of the definition of the GTES $P_i$, $i\geq 1$.
\begin{lem}\label{telekP}
(a) For each $i\geq 1$,  
 $\tthue {P_i}=\tthue {R_i}  \subseteq \tthue {P_{i+1}}\subseteq \tthue E$.

 (b) If $R_i=P_{i+1}$ for some $i\geq 1$, then 
$R_i=P_j=R_j$ for  $j\geq i+1$. 
\end{lem}


\begin{lem} \label{zsur}For each $i\geq 1$, 
we can effectively construct the GTES $P_i$.
\end{lem}
\pr By induction on $i$.

{\em Base Case:} $i=1$. Clearly, we can construct  $P_1$.

{\em Induction Step:} Let $i\geq 1$.  Assume that we have constructed
$P_i$.  
By Proposition \ref{kairo}, 
we can construct  $R_i$. Consider item (b) in the definition of $P_i$. 
By Proposition \ref{eldont}, 
we can effectively decide whether 
$R_i$  reaches $p\hspace{-1mm}\downarrow_{R_i}$ 
starting from $l[u_1 \ldots, u_m]\hspace{-1mm}\downarrow_{R_i}$ attached to some context.
Hence we can construct $P_{i+1}$ as well. 

\hfill $\b$


We now present our semi-decision procedure. 

\vspace{ 3mm} 

\noindent
{\bf Procedure } $\textit{PRO3}$
{\em Input:} A  variable preserving TES $E$ over the ranked alphabet $\S$ and ground terms $p, q\in \ts$.

\noindent
{\em Output:} 
$\bullet$ 'yes'  if   $p\tthue E q$,

 \hspace{7.2 mm} $\bullet$ 'no' if  $(p, q)\not\in\tthue E $  and   the procedure halts,

\hspace{7.2 mm} $\bullet$ undefined if the procedure does not halt.

compute $P_1$, $R_{P_1}$, $Q_1$, $R_{Q_1}$, and   $R_{P_1\cup Q_1}$;

if $p\hspace{-1mm}\downarrow_ {R_{P_1\cup Q_1}}= q\hspace{-1mm}\downarrow_ {R_{P_1\cup Q_1}}$,  then begin output 'yes'; halt end;
  
$i:=1$;

1: $i:=i+1$;

compute
 $P_i$, $R_{P_i}$, $Q_i$, $R_{Q_i}$, and  $R_{P_i\cup Q_i}$;
 
if $p\hspace{-1mm}\downarrow_ {R_{P_i \cup Q_i}}= q\hspace{-1mm}\downarrow_ {R_{P_i \cup Q_i}}$,   then begin output 'yes'; halt end;

if $R_{P_{i-1}}=P_i$ or  
  $R_{Q_{i-1}}=Q_i$, 

 then 
begin output 'no'; halt end;

 goto 1

\begin{exa}\label{korte}{\em 
We continue  Example \ref{takacs}. Note that 
$p\hspace{-1mm}\downarrow_{R_{P_1\cup Q_1}}\neq 
q\hspace{-1mm}\downarrow_{R_{P_1\cup Q_1}}$.
 Hence procedure $\textit{PRO3}$ does not output anything and does not 
halt in the first  step. 
Observe that  
$p\hspace{-1mm}\downarrow_{R_{P_2\cup Q_2}}=
q\hspace{-1mm}\downarrow_{R_{P_2\cup Q_2}}$.
Hence 
procedure $\textit{PRO3}$   outputs 'yes' and  halts in the second  step.

\hfill $\b$ }
\end{exa}

\begin{exa}\label{dalol} {\em We continue Example \ref{cipo}. Let 
$n\geq 1$.
We run    procedure $\textit{PRO3}$
on   the  TES $E$ and the ground terms
$p=(bd)^{2n}\$$, and $q=(db)^{2n}\$$. We    compute as follows.
GTES $P_1$  consists of the equation  $bd\$\approx db\$$.
Let $\Theta$ stand for $\tthue {P_1}\cap (sub(P_1) \times sub(P_1))$.
Then 
$sub(P_1)/\Theta=\{\, \{\, b\$\, \}, \{\, d\$\, \}, \{\, bd\$\, \}\, \}$ and $\{\, bd\$ \, \}$ 
is a set of representatives for $sub(P_1)/\tthue {P_1}$.
GTRS $R_{P_1}$ consists  of the rule
$bd\$\rightarrow db\$$.

Symmetrically, GTES $Q_1$  consists of the equation  $db\$\approx bd\$$.
GTRS $R_{Q_1}$ consists  of the rule
$db\$\rightarrow bd\$$. It is not hard to see, that 
GTRS $R_{P_1\cup Q_1}$ is equal to GTRS $R_{P_1}$.
Observe that 
 $p\hspace{-1mm}\downarrow_ {R_{P_1\cup Q_1}}= q\hspace{-1mm}\downarrow_ {R_{P_1\cup Q_1}}$,
Hence  procedure $\textit{PRO3}$ outputs 'yes' and halts in the first step.

We run    procedure $\textit{PRO3}$
on   the  TES $E$ and the ground terms
$p=aaa\$$ and $q=bedb\$$. By our arguments in Example \ref{cipo}, 
$$p\hspace{-1mm}\downarrow_ {R_{P_i \cup Q_i}}\neq  
q\hspace{-1mm}\downarrow_ {R_{P_i \cup Q_i}}\mbox{ for } i\geq 1\, .$$ 
  Furthermore, $\textit{PRO3}$ computes as follows.

$R_{Q_1}=\{\,db\$ \rightarrow bd\$,   edb\$ \rightarrow de\$\, \}$, 

$R_{Q_2}=\{\, db\$ \rightarrow bd\$,   edb\$ \rightarrow de\$,  bdde\$ \rightarrow dbde\$  \, \}$, and

$R_{Q_2}=R_{Q_{n+2}}$ for $n\geq 1$.

\noindent
Consequently,  Procedure 
$\textit{PRO3}$ outputs 'no' and then  halts. 
Generalizing our arguments, we can show the following.

\begin{sta}
Let $p\in \ts$ be arbitrary such that  symbols $a$ or $c$ appear in $p$.
  Let $q\in \ts$ such that $a, c$ do not appear in $q$.
Then  procedure $\textit{PRO3}$ outputs 'no' and halts
on the input $E$, $p$, $q$. 
\end{sta}


By Propositon \ref{halasz},  for an arbitrary ground term $q'\in \ts$,
the  goal-directed completion procedure \cite{lyn} may fail or may not  halt  on  the TES $E$ and the goal $(aaa\$, q')$.
The following problem is open.
For each  goal  $(aaa\$, q)$
such that $q\in \ts$, and $a, c$ do not appear in $q$, is it true that the the goal-directed completion procedure
does not fail and 
halts on the TES $E$ and the  goal  $(aaa\$, q)$.

It is  open whether  the goal-directed completion procedure does not fail and halts on the TES $E$ and any goal  $(aaa\$, q)$
such that $q\in \ts$, $a, c$ do not appear in $q$.

\hfill $\b$
}
\end{exa}

We now show the correctness of Procedure $\textit{PRO3}$. 
\begin{lem}\label{roma}
For any $i, n$ with 
$1\leq n\leq i$, and any  $t_1, \ldots, t_n \in \ts$, if
$p\thue E t_1 \thue E t_2 \thue E \cdots \thue E t_n$,
then

\noindent 
$p\tthue {P_i} t_1 \tthue  {P_i} t_2 \tthue  {P_i} \cdots \tthue  {P_i} t_n$.

\end{lem}
\pr We proceed by induction on $i$.

{\em Base Case}: $i=1$. Then $n=1$. By the definition of $P_1$, we have 
$p\thue {P_1} t_1$. 

{\em Induction Step}: Let $i\geq 1$, and assume that the statement holds for $1, 2, \ldots, i$.
We now show that the  statement holds for $i+1$.
To this end, assume that 
\begin{equation}\label{kosar}
p\thue E t_1 \thue E t_2 \thue E \cdots \thue E t_{n} \mbox{ for some }
0\leq n\leq i+1\,.
\end{equation}
By the induction  hypothesis,
\begin{equation}\label{dallam}
p\tthue {P_i} t_1 \tthue  {P_i} t_2 \tthue  {P_i} \cdots \tthue  {P_i} t_{n-1}\, .
\end{equation}
Hence
\begin{equation}\label{rem}
t_{n-1} \tred {R_i} p\hspace{-1mm} \downarrow_{R_i} \, . 
\end{equation}
By (\ref{kosar}),  there is an equation
 $l\approx r$ in $E\cup E^{-1}$ with $l, r \in \ts(X_m)$, $m\geq 0$
and there are   $u\in C_\S$,  $u_1, \ldots, u_m\in\ts$ such that 
\begin{equation}\label{zalam}
t_{n-1}=u[l[u_1, \ldots, u_m]] \mbox{ and }
t_{n}=u[r[u_1, \ldots, u_m]]\,.
\end{equation} 
As $R_i$ is convergent, by (\ref{rem}) and (\ref{zalam}), 
$u[l[u_1, \ldots, u_m]\hspace{-1mm} \downarrow_{R_i}] \tred {R_i} p\hspace{-1mm} \downarrow_{R_i}$. That is,
$R_i$  reaches $p\hspace{-1mm}\downarrow_{R_i}$ starting from $l[u_1, \ldots, u_{m}]\hspace{-1mm}\downarrow_{R_i}$ attached to some context.
By the definition of $P_{i+1}$, 
\begin{equation}\label{30y}
l[u_1, \ldots, u_m] \approx r[u_1, \ldots, u_m]
\mbox{ is in }\tthue {P_i} \mbox{  or } P_{i+1}\, .
\end{equation}
By  Lemma \ref{telekP},  (\ref{dallam}), (\ref{zalam}), and (\ref{30y}), 
$$p\tthue {P_{i+1}} t_1 \tthue  {P_{i+1}} t_2 \tthue  {P_{i+1}} \cdots \tthue  {P_{i+1}} t_{n-1}
\tthue  {P_{i+1}} t_{n}\, .$$

\hfill $\b$

\noindent
By Lemma \ref{telekP}  and Lemma \ref{roma} we have the following result.
\begin{lem}\label{osz} Assume that 
$R_i=P_{i+1}$ for some $i\geq 1$.  Then 
$p \tthue {P_{i+1}} q$ if and only if $p \tthue E q$. 

\end{lem}

\begin {tet}\label{egy} If $p\tthue E q$, then procedure $\textit{PRO3}$  outputs 'yes' and  halts. 
\end{tet}
\pr  
Assume that $p= t_1 \thue E t_2 \thue E \cdots \thue E t_n=q$ for some 
$n\geq 1$ and $t_1, \ldots, t_n \in \ts$.
By Lemma \ref{roma}, 
$p\tthue {P_n} q $.
Let $k$ be the least integer such that $p\tthue {P_k\cup Q_k} q$. 

First assume that $k=1$. Then  $p \tthue {P_1\cup Q_1} q$.   Hence 
$p\hspace{-1mm}\downarrow_ {R_{P_1\cup Q_1}}= q\hspace{-1mm}\downarrow_ {R_{P_1\cup Q_1}}$.
Consequently,
 procedure $\textit{PRO3}$  outputs 'yes' and  halts in the first step.

Second assume that  $k\geq 2$. 
Then by the definition of $k$, $(p, q)\not\in\tthue {P_i\cup Q_i}$   for  $2\leq i \leq k-1$.  
Then by Lemma  \ref{osz}, $R_{P_{i-1}}\subset P_i$  and $R_{Q_{i-1}}\subset Q_i$  for  $2\leq i \leq k-1$.  
Hence procedure $\textit{PRO3}$  does not halt in the first $k-1$    steps. 
By the definition of the integer $k$, 
in the $k$th      step  procedure $\textit{PRO3}$  outputs 'yes'  and  halts.

\hfill $\b$

\begin {tet} \label{ketto} If  procedure $\textit{PRO3}$  outputs 'yes'  and  halts, then 
$p \tthue E q$. If procedure $\textit{PRO3}$  outputs 'no' and  halts,  then 
$(p , q) \not \in \tthue E $.

\end{tet}
\pr Assume that procedure $\textit{PRO3}$  outputs 'yes'  and  halts in the $k$th    step. 
Then   $p\tthue {P_k\cup Q_k} q$. By Lemma \ref{telekP}, $p \tthue E q$.

Assume that procedure $\textit{PRO3}$  outputs 'no'   and  halts in the $k$th    step. 
Then 

(a) $(p, q) \not\in  \tthue {P_k\cup Q_k}$ and 

(b) $P_k=R_{P_{k-1}}$ or  $Q_k=R_{Q_{k-1}}$. 

\noindent
We now distinguish two cases.

{\em Case 1:} $P_k=R_{P_{k-1}}$. By (a) 
and by Lemma  \ref{osz}, $(p, q) \not \in \tthue E$.

{\em Case 2:} $Q_k=R_{Q_{k-1}}$.
This case is symmetric to Case 2.

\hfill $\b$

Theorems \ref{egy} and \ref{ketto} imply the following.
\begin {tet} \label{fotetel} If   
$p \tthue E q$, then procedure $\textit{PRO3}$  outputs 'yes'  and  halts. Otherwise,  either $\textit{PRO3}$  outputs 'no' and  halts, or 
 $\textit{PRO3}$ does not halt. 

\end{tet}

\begin{exa}
{\em We continue Example \ref{kiraly}.
We now run  procedure $\textit{PRO3}$
 on   the TES $E$ and the ground terms  $p$, $q$. Then 
$P_1=Q_1=\{\,   \# \approx \$\, \}$, 
 $R_{P_1}=R_{Q_1}=P_1$, and  $R_{P_1\cup Q_1}=P_1$. 
 Observe that $p\hspace{-1mm}\downarrow_ {R_{P_1\cup Q_1}}= q\hspace{-1mm}\downarrow_ {R_{P_1\cup Q_1}}$. 
 Hence procedure
  $\textit{PRO3}$    outputs 'yes' and  halts in the first step. 
By Proposition \ref{bartok},  we compute $p\hspace{-1mm}\downarrow_{R_{P_1\cup Q_1}}$ and 
$q\hspace{-1mm}\downarrow_{R_{P_1\cup Q_1}}$ in linear time. We apply the rules of ${R_{P_1\cup Q_1}}$ $n$ times.
 For this example,  $\textit{PRO3}$ is faster than $\textit{PRO1}$.

\hfill $\b$ }
\end{exa}


\begin{exa}{\em  We continue Example \ref{tavasz}.
We now run procedure $\textit{PRO3}$
 on the   TES $E$ and the ground terms  $p$ and $q$.
Then 
$\{\, a \approx  ffa\, \}=P_1=R_{P_1}=P_{1+i}=R_{P_{1+i}}$ for $i\geq 1$. 
Furthermore, 
$Q_1=\{\,  a \approx   ffa, \, fa \approx  fffa\, \, \}$, $R_{Q_1}=P_1=Q_2 = R_{Q_2}=
Q_{1+i}=R_{Q_{1+i}}$ for $i\geq 1$. 

Observe that 
$p\hspace{-1mm}\downarrow_{R_{P_2\cup Q_2}}\neq 
q\hspace{-1mm}\downarrow_{R_{P_2\cup Q_2}}$.
 Hence procedure
  $\textit{PRO3}$    outputs 'no' and  halts in the second step. 
   
It should be clear that for all ground terms $p$ and $q$,  $\textit{PRO3}$ halts. It  outputs 'yes'   if 
$p \tthue E q$. Otherwise it  outputs 'no'.

\hfill $\b$ 
}
\end{exa}

\begin{exa}  \label{uszo} {\em
We now continue Example \ref{futo}.
We apply procedure $\textit{PRO3}$ to the TES $E=\{\, ffx\approx gfx\, \}$ and any terms $p, q\in \ts$. 
 Observe that $height(ffx)=2=height(gfx)$. 


\begin{sta}\label{uhu}
For each $i\geq 0$, and for  each pair of terms, $s, t \in \ts(X)$, if $(s, t) \in P_i$, then 
 $height(s)=height(t)\leq height(p)$. 
\end{sta}
\pr
We proceed by induction on $n$.

{\em Base Case}: $i=1$.
By the definition of $P_1$, for  each equation $s \approx t$ in $P_1$, $height(s)=height(t)\leq height(p)$.  Hence our statement  holds.

{\em Induction Step}: Let $n\geq 1$, and assume that the satement holds for $1, 2, \ldots, n$.
We now show that the  satement holds for $n+1$.
Consider an equation  
$l[u_1, \ldots, u_m] \approx r[u_1, \ldots, u_m]$
in $P_{i+1}$. Then there exist

$\bullet$ an   equation $l\approx r$ of $E\cup E^{-1}$, where $l, r \in\ts(X_m)$, $m\geq 0$. 

$\bullet$ 
$u_1, \ldots,  u_m\in (sub(R_i)-lhs(R_i))\cup sub( p\hspace{-1mm}\downarrow_{R_i})$. 

\noindent
such that 
$R_i$  reaches $p\hspace{-1mm}\downarrow_{R_i}$ starting from $l[u_1,  \ldots, u_m]\hspace{-1mm}\downarrow_{R_i}$ attached to some context, and
that

\noindent
$l[u_1, \ldots, u_m]\hspace{-1mm}\downarrow_{R_i} \neq  r[u_1, \ldots, u_m]\hspace{-1mm}\downarrow_{R_i} $.

\noindent Consequently, there is a $u\in C_\S$ 
such that $u[l[u_1,  \ldots, u_m]]\tred {R_i} p$. By (a) in Lemma \ref{telekP} and the induction hypothesis,
  $height(u[l[u_1,  \ldots, u_m]]) =height(p)$.
Thus $height(l[u_1, \ldots, u_m] ) \leq height(p)$.  By (a) in Lemma \ref{telekP} and the induction hypothesis,
$height(l)=height(r)$. Hence 
$height(l[u_1,  \ldots, u_m])= height(r[u_1,  \ldots, u_m])$.

\hfill $\b$


Observe that the set $\{\, (s, t)\in \ts\times \ts\mid height(s)=height(t)\leq height(p)\, \}$ is finite.
By  Lemma \ref{telekP} and Statement \ref{uhu},  procedure $\textit{PRO3}$ halts on $E$  and any terms $p, q\in \ts$
in finitely many steps.

\hfill $\b$ }
\end{exa}

The following result can be shown by generalizing the proof appearing in Example \ref{uszo}. 
\begin{tet}
Let $E$ be a  variable preserving TES
such that

$\bullet$ for any equation $s\approx t$ in $E$, $height(s)=height(t)$, or 

$\bullet$ for any equation $s\approx t$ in $E$, $size(s)=size(t)$ and each variable appears  the same times in $s$ and $t$.

\noindent Let $p, q\in \ts$ be arbitrary.
Then  procedure $\textit{PRO3}$ halts on $E$  and terms $p, q$.
\end{tet}


\section{Semi-decision procedure for the ground
             word problem of TESs}\label{2fo}
We present the semi-decision procedure $\textit{PRO4}$ for the ground
 word problem of TESs,   and show its correctness.
 We obtain it  generalizing  $\textit{PRO3}$ taking into account  $\textit{PRO2}$. The starting point to the definition of 
 the GTESs $P_i$, $i\geq 1$,  is the same as in Section \ref{fo}. 
 We define $P_1$ as the set of all  instances $l'\rightarrow r'$ of equations  $l\approx r$
 in $E\cup E^{-1}$ which 
can be applied to $p$. 
 We define $P_{i+1}$, $i\geq 1$,  as the set of all  instances $l'\rightarrow r'$ of equations  $l\approx r$
 in $E\cup E^{-1}$ which 
can be applied to elements of $\{\, s\in \ts \mid p \tthue {P_i} s\, \}$.
The question is what should we substitute for those  variables in the right-hand side $r$ 
that do not appear in the left-hand side $l$. We now give a  simplified answer to this question. 
Applying Snyder's algorithm   we compute a reduced GTRS $R_i$ equivalent to the GTES $P_i$. 
When constructing the instance  $l'\rightarrow r'$ of  $l\approx r$, 
 we substitute any term in $(sub(R_i)-lhs(R_i))\cup sub( p\hspace{-1mm}\downarrow_{R_i})$
or the $R_i$ normal form of any  ground term of height less than or equal to $i$
for each  variable in the right-hand side $r$ that does not appear on the left-hand side $l$. 
Furthermore,  
we define the GTESs $Q_i$, $i\geq 1$,  symmetrically.

Let $E$ be a TES over $\S$, and let $p, q\in \ts$. 
We now define the GTESs $P_i$  and the reduced GTRSs $R_i$,
$i \geq 1$, 
 over $\S$. 

Let $NORM_0=\S_0\cup sub(p)$.  
 For each equation  $l\approx r$ of $E\cup E^{-1}$ 
with  $l\in T_\S(X_{k+m})$, $r \in T_\S(X_k\cup X_{[k+m+1,  k+m+\ell]})$ for some 
$k, m, \ell\geq 0$, 
if  
$p=u[l[u_1, \ldots, u_{k+m}]]$  for some  $u\in C_\S$,
$u_1, \ldots, u_{k+m}\in\ts$, 
then for all $v_{k+m+1}, \ldots, v_{k+m+\ell}\in NORM_0$, we put the  equation
$$l[u_1, \ldots, u_{k+m}] \approx r[u_1, \ldots, u_k, v_{k+m+1}, \ldots, v_{k+m+\ell}]$$
in $P_1$. 
Applying Snyder's algorithm   we compute a reduced GTRS $R_1$ equivalent to the GTES 
$P_1$, see Proposition \ref{kairo}.

Let $i\geq 1$. 
Let 

$NORM_i=sub(p\hspace{-1mm}\downarrow_{R_i} ) \cup (sub(R_i)-lhs(R_i)) \cup $

$
\{\, t\hspace{-1mm}\downarrow_{R_i} \mid t\in  NORM_{i-1} \mbox{ or }
t=f(t_1, \ldots, t_m)  \mbox{ for some }f\in \S_m  \mbox{ and } t_1, \ldots, t_m \in NORM_{i-1}
\, \}$. 

(a)
We put 
each rule  of $R_i$ into $P_{i+1}$.

(b) For each equation  $l\approx r$ of $E\cup E^{-1}$ 
with  $l\in T_\S(X_{k+m})$, $r \in T_\S(X_k\cup X_{[k+m+1,  k+m+\ell]})$ for some 
$k, m, \ell\geq 0$, 
for any $u_1, \ldots,  u_{k+m}\in (sub(R_i)-lhs(R_i))\cup sub( p\hspace{-1mm}\downarrow_{R_i})$ and 
$v_{k+m+1}$, $\ldots$,  $v_{k+m+\ell}\in NORM_i$,
if 
$R_i$  reaches $p\hspace{-1mm}\downarrow_{R_i}$ starting from $l[u_1, \ldots, u_{k+m}]\downarrow_{R_i}$ attached to some context, 
 and

$l[u_1, \ldots, u_{k+m}]\hspace{-1mm}\downarrow_{R_i} \neq  
r[u_1, \ldots, u_m,v_{k+m+1}, \ldots, v_{k+m+\ell} ]\hspace{-1mm}\downarrow_{R_i} $,

\noindent then we put the equation  
$$l[u_1, \ldots, u_{k+m}] \approx r[u_1, \ldots, u_m,v_{k+m+1}, \ldots, v_{k+m+\ell} ]$$
in $P_{i+1}$.

If we do not put equations in $P_{i+1}$ in item (b), i.e. $P_{i+1}=R_i$, then let $R_{i+1}=R_i$. Otherwise, 
applying Snyder's algorithm,   we compute a reduced GTRS $R_{i+1}$ equivalent to the GTES $P_{i+1}$.

When misunderstanding may arise, we denote $R_i$ as  $R_{P_i}$.
We define the GTESs $Q_i$, $i\geq 1$, symmetrically to the GTESs $P_i$, $i\geq 1$.
Applying Snyder's algorithm,   we compute a reduced GTRS $R_{P_i \cup Q_i}$ equivalent to the GTRS
$R_{P_i} \cup R_{Q_i}$ for $i\geq 1$. 

By Proposition \ref{mozart}  GTRSs $R_{P_i}$,  $R_{Q_i}$, and  $R_{P_i \cup Q_i}$ are convergent.

We illustrate our concepts and results by two running examples, each of them  is presented as a 
series of examples. 
\begin{exa}\label{2spenot}{\em We continue  Example \ref{makaroni}. 
Let $p=f(0, 1)$ and $q=f(f(0, 1), 1)$. 
Observe that for any $u, v \in \ts$, if $u \tthue E v$, then the parity of the number of $1$'s  in $u$ equals to that in  $v$. 
Hence 
\begin{equation}\label{agrar}
(p, q)\not \in \tthue E\, . 
\end{equation}

We now construct the GTESs $P_1$, $P_2$,  and $P_3$. 
Then $NORM_0=\{\, 0, 1, f(0, 1)\, \}$.  
$P_1$ consists of the  equations

$0 \approx  f(0, 0)$, \hspace{3 mm}
$ 0 \approx f(1,1)$, \hspace{3 mm}
$ 0 \approx f(f(0, 1),f(0, 1))$, \hspace{3 mm}

$ 1\approx f(0, 1)$,  \hspace{3 mm}
$ f(0, 1)\approx 1$,  \hspace{3 mm}
$f(0, 1)\approx f(0,f(0, 1))$.
 
\noindent $R_1$  consists of the  rules

$f(0, 0)\rightarrow   0$, \hspace{3 mm}
$f(1,1)\rightarrow   0$, \hspace{3 mm}
$f(0, 1)\rightarrow   1$.

\noindent $NORM_1 =\{\, 0, 1, f(1, 0)\, \}$.
$P_2$ consists of the  equations

$f(0, 0)\approx  0$, \hspace{3 mm}
$f(1,1)\approx 0$, \hspace{3 mm}
$f(0, 1)\approx  1$, \hspace{3 mm}
$0\approx f(f(1, 0), f(1, 0))$.

\noindent $R_2$  consists of the  rules

$f(0, 0)\rightarrow   0$, \hspace{3 mm}
$f(1,1)\rightarrow   0$, \hspace{3 mm}
$f(0, 1)\rightarrow   1$,  \hspace{3 mm}
$f(f(1, 0), f(1, 0))\rightarrow  0$. 

\noindent $NORM_2 =\{\, 0, \; 1, \; f(1, 0), \; f(0, f(1, 0)), \;
f(1, f(1, 0)), \; f(f(1, 0), 0), \; f(f(1, 0), 1)
\, \}$.

\noindent $P_3$ consists of the  equations

$f(0, 0)\approx  0$, \hspace{3 mm}
$f(1,1)\approx 0$, \hspace{3 mm}
$f(0, 1)\approx  1$, \hspace{3 mm}
$f(f(1, 0), f(1, 0))\approx  0$, 

$0\approx  f( f(0, f(1, 0)) ,f(0, f(1, 0)) )$,

$0\approx  f( f(1, f(1, 0))  ,  f(1, f(1, 0))  )$, 

$0\approx  f(f(f(1, 0), 0)   ,  f(f(1, 0), 0) )$, 

$0\approx  f( f(f(1, 0), 1) , f(f(1, 0), 1)  )$. 

\noindent $R_3$  consists of the  rules

$f(0, 0)\rightarrow   0$, \hspace{3 mm}
$f(1,1)\rightarrow   0$, \hspace{3 mm}
$f(0, 1)\rightarrow   1$,  \hspace{3 mm}
$f(f(1, 0), f(1, 0))\rightarrow  0$, 

$  f( f(0, f(1, 0)) ,f(0, f(1, 0)) ) \rightarrow   0$, 

$f( f(1, f(1, 0))  ,  f(1, f(1, 0))  )  \rightarrow   0$, 

$  f(f(f(1, 0), 0)   ,  f(f(1, 0), 0) ) \rightarrow   0$, 

$f( f(f(1, 0), 1) , f(f(1, 0), 1)  )  \rightarrow   0$. 

\noindent 
Continuing in this manner we get that 
\begin{equation}\label{buza}
R_{P_{i}}\subset R_{P_{i+1}}\mbox{  for } i\geq 1\, .
\end{equation}

We now compute the  GTESs $Q_1$,  $Q_2$, and $Q_3$. 

\noindent $NORM_0=\{\, 0, 1, f(0, 1), f(f(0, 1), 1)\, \}$. 

\noindent $Q_1$ consists of the  equations

$0 \approx f(0, 0)$, \hspace{3 mm}
$0 \approx f(1,1)$, \hspace{3 mm}
$0 \approx f(f(0, 1),f(0, 1))$, \hspace{3 mm}
$0 \approx f( f(f(0, 1), 1), f(f(0, 1), 1))$.

$1 \approx f(0, 1)$,  \hspace{3 mm}
$f(0, 1) \approx f(0,f(0, 1)) $,  \hspace{3 mm}
$f(f(0, 1), 1) \approx f(0, f(f(0, 1), 1))$. 

\noindent $R_{Q_1}$  consists of the  rules

$f(0, 0)\rightarrow 0$, \hspace{3 mm}
$f(1,1)\rightarrow 0$, \hspace{3 mm}
$f(0, 1)\rightarrow 1$.

\noindent $NORM_1 =\{\, 0, 1, f(1, 0)\, \}$.

\noindent $Q_2$  consists of the  equations

$f(0, 0)\approx 0$, \hspace{3 mm}
$f(1,1)\approx 0$, \hspace{3 mm}
$f(0, 1)\approx 1$, \hspace{3 mm}
$0 \approx f(f(1, 0), f(1, 0))$.

\noindent $R_{Q_2}$  consists of the  rules

$f(0, 0)\rightarrow 0$, \hspace{3 mm}
$f(1,1)\rightarrow 0$, \hspace{3 mm}
$f(0, 1)\rightarrow 1$, \hspace{3 mm}
$f(f(1, 0), f(1, 0))\rightarrow 0$.

$NORM_2=\{\, 0, 1, f(1, 0), f( 0,  f(1, 0)), f(1,  f(1, 0)), f( f(1, 0), 0), f( f(1, 0), 1)
\, \}$.

\noindent $Q_3$  consists of the  equations

$f(0, 0)\approx 0$, \hspace{3 mm}
$f(1,1)\approx 0$, \hspace{3 mm}
$f(0, 1)\approx 1$, \hspace{3 mm}
$0 \approx f(f(1, 0), f(1, 0))$,

 $0 \approx f(  f( 0,  f(1, 0))  , f( 0,  f(1, 0))  )$,   \hspace{3 mm}
$0 \approx f(  f(1,  f(1, 0)), f(1,  f(1, 0)) )$, 

$0 \approx f(  f( f(1, 0), 0)  ,  f( f(1, 0), 0)  )$, \hspace{3 mm}
$0 \approx f(  f( f(1, 0), 1)  , f( f(1, 0), 1) )$.

\noindent $R_{Q_3}$  consists of the  rules

$f(0, 0)\rightarrow 0$, \hspace{3 mm}
$f(1,1)\rightarrow 0$, \hspace{3 mm}
$f(0, 1)\rightarrow 1$, \hspace{3 mm}
$f(f(1, 0), f(1, 0))\rightarrow 0$,

 $f(  f( 0,  f(1, 0))  , f( 0,  f(1, 0))  )\rightarrow 0$,   \hspace{3 mm}
$f(  f(1,  f(1, 0)), f(1,  f(1, 0)) )\rightarrow 0$, 

$f(  f( f(1, 0), 0)  ,  f( f(1, 0), 0)  )\rightarrow 0$, \hspace{3 mm}
$f(  f( f(1, 0), 1)  , f( f(1, 0), 1) )\rightarrow 0$.

\noindent 
Continuing in this manner we get that 
\begin{equation}\label{arpa}
R_{Q_{i}}\subset R_{Q_{i+1}}\mbox{  for } i\geq 1\, .
\end{equation}

\noindent Let 
$R_{P_1\cup Q_1}= R_{P_1}$, $R_{P_2\cup Q_2}= R_{P_2}$, and 
 $R_{P_3\cup Q_3}= R_{P_3} \cup R_{Q_3}$. 

\hfill $\b$ 
 }
\end{exa}


\begin{exa}\label{repa2}{\em 
Let $\S=\S_0\cup \S_1$, $\S_0=\{\, 0, 1\, \}$, and $\S_1=\{\, g, h\, \}$.
Let TES $E$ consist of the  equations

$gx_1\approx x_1$, \hspace{3 mm} 
$hx_1\approx  hx_2$.

 \noindent Let $p=0$ and $q=1$. 

We now construct the GTESs $P_1$, $P_2$, and $P_3$. 
Then $NORM_0=\{\, 0, 1\, \}$.  
$P_1$ consists of the  equation
$0\approx  g0$. \hspace{3 mm}

\noindent 
 $R_1$ consists of the rule $g0\rightarrow 0$. 

\noindent 
 $NORM_1=\{\, 0, 1, g1, h0, h1\, \}$.  

\noindent 
$P_2=R_1$ and 
$R_2=P_2$.
 
\noindent 
$NORM_2=\{\, 0, 1, g1, h0, h1, gg1, hg1, gh0, hh0, gh1, hh1
\, \}$.  

\noindent 
$P_3=R_2$ and 
$R_3=P_3$.

We now construct the GTESs $Q_1$, $Q_2$,   and $Q_3$. 
Then $NORM_0=\{\, 0, 1\, \}$.  
$Q_1$ consists of the  equation $1\approx  g1$. 

\noindent 
$R_{Q_1}$ consists of the rule $g1\rightarrow 1$.

\noindent $NORM_1=\{\, 0, 1, g0, h0, h1
\, \}$.  

\noindent 
$Q_2=R_{Q_1} $ and 
$R_{Q_2}=Q_2$.

\noindent 
$NORM_2=\{\, 0, 1, g0, h0, h1, 
gg0, hg0, gh0, hh0, gh1, hh1
\, \}$.  

\noindent 
$Q_3=R_{Q_2} $ and 
$R_{Q_3}=Q_3$.

\noindent $R_{P_1}\cup R_{Q_1}= R_{P_1\cup Q_1}= R_{P_2\cup Q_2}=R_{P_3\cup Q_3}$.
 
\hfill $\b$ }
\end{exa}


We get the following result
by direct inspection of the definition of the  GTES $P_i$ and  GTRS $R_i$, $i\geq 1$.
\begin{sta} \label{2babyP} For each $i\geq 1$, 
$\tthue {P_i}\subseteq \tthue {P_{i+1}}\subseteq \tthue E$. 
\end{sta}
We can show the following result similarly to Lemma \ref{zsur}. 
\begin{lem} \label{2zsur}For each $i\geq 1$, 
we can effectively construct the GTES $P_i$.
\end{lem}


\begin{lem} \label{talalkozas} For each  $i\geq 1$, 
$ sub(p\hspace{-1mm}\downarrow_{R_{P_i}})\cup
(sub(R_{P_i})-lhs(R_{P_i})) \cup 
\{\, t\hspace{-1mm}\downarrow_{R_{P_i}} \mid height(t) \leq i\, \}
\subseteq NORM_i$.
\end{lem}
\pr By induction on $i$.

\hfill $\b$

We now present our semi-decision procedure. 

\vspace{ 3mm} 

\noindent
{\bf Procedure } $\textit{PRO4}$
{\em Input:} A  variable preserving TES $E$ over the ranked alphabet $\S$ and ground terms $p, q\in \ts$.

\noindent
{\em Output:} 
$\bullet$ 'yes'  if   $p\tthue E q$,

 $\bullet$ 'no' if  $(p, q)\not\in\tthue E $  and   the procedure halts,

$\bullet$ undefined if the procedure does not halt.

compute $P_1$, $R_{P_1}$, $Q_1$, $R_{Q_1}$, and   $R_{P_1\cup Q_1}$;

if $p\hspace{-1mm}\downarrow_ {R_{P_1\cup Q_1}}= q\hspace{-1mm}\downarrow_ {R_{P_1\cup Q_1}}$,  then begin output 'yes'; halt end;
  
$i:=1$;

1: $i:=i+1$;

compute
 $P_i$, $R_{P_i}$, $Q_i$, $R_{Q_i}$, and  $R_{P_i\cup Q_i}$;
 
if $p\hspace{-1mm}\downarrow_ {R_{P_i \cup Q_i}}= q\hspace{-1mm}\downarrow_ {R_{P_i \cup Q_i}}$,   then begin output 'yes'; halt end;
 
if $i=2$, then goto 1;
   
if $R_{P_{i-2}}=R_{P_{i-1}}=P_{i}$,  or  $R_{Q_{i-2}}=R_{Q_{i-1}}=Q_{i}$,

 then 
begin output 'no'; halt end;

 goto 1

\begin{exa}
{\em  We continue Example \ref{2spenot}. 
By Statement \ref{2babyP}   and (\ref{agrar}),
$p\hspace{-1mm}\downarrow_{R_{P_i\cup Q_i}}\neq
q\hspace{-1mm}\downarrow_{R_{P_i\cup Q_i}}$ for $i\geq 1$.  Hence procedure 
$\textit{PRO4}$ does not output 'yes'.
By (\ref{buza}) and (\ref{arpa}), 
procedure 
$\textit{PRO4}$ does not output 'no'.
Hence procedure $\textit{PRO4}$ does not output anything and does not 
halt at all. 

}
\hfill $\b$ 
\end{exa}

\begin{exa}{\em 
We continue Example \ref{repa2}. 
Observe  that 

$p\hspace{-1mm}\downarrow_{R_{P_1\cup Q_1}}=0\neq 1=
q\hspace{-1mm}\downarrow_{R_{P_1\cup Q_1}}$,

 $p\hspace{-1mm}\downarrow_{R_{P_2\cup Q_2}}=0\neq 1=
q\hspace{-1mm}\downarrow_{R_{P_2\cup Q_2}}$, 

$p\hspace{-1mm}\downarrow_{R_{P_3\cup Q_3}}=0\neq 1=
q\hspace{-1mm}\downarrow_{R_{P_3\cup Q_3}}$, and 

$R_{P_{1}}=R_{P_{2}}=P_{3}$. 

\noindent
Hence procedure $\textit{PRO4}$ outputs 'no' and   halts in the third   step.

\hfill $\b$}
\end{exa}

\begin{exa}\label{sizo} {\em  Let  $\S=\S_0\cup \S_1\cup \S_2$, $\S_0=\{\,\$, \#\, \}$, $\S_1=\{\,f, g\, \}$,
$\S_2=\{\,h\, \}$.
Consider the TES $E=\{\, ffx_1\approx gfx_1, \; h(x_1, x_1)\approx \$\, \}$. 
As in Example \ref{futo}, we can show that the basic  Knuth-Bendix completion procedure runs forever on 
this example. Moreover, it is still open
 whether  the goal-directed completion procedure halts on the TES $E$ and any goal.

Let $n\geq 1$.  Let $p= h(f^n\$, gf^{n-1}\$)$ and $q= \$$. We raise the problem whether
$ p \tthue E q$. 
We now apply procedure $\textit{PRO4}$ to the TES $E$ and the  terms $p, q$. 

\noindent
GTRS $R_{P_1}$ consists of the rules

 $f^i\$ \rightarrow  gf^{i-1}\$$ for $2\leq i \leq n$, 

$h(\$, \$)\rightarrow \$$, 

$h(\#, \#)\rightarrow \$$.  

\noindent
GTRS $R_{Q_1}$ consists of the rules

$h(\$, \$)\rightarrow \$$, 

$h(\#, \#)\rightarrow \$$.  

\noindent
GTRS $R_{P_2}$ consists of the rules

 $f^2gf\$ \rightarrow  gf\$$, 

$h(\$, \$)\rightarrow \$$, 

$h(\#, \#)\rightarrow \$$.

$h(f\$, f\$)\rightarrow \$$, 

$h(f\#, f\#)\rightarrow \$$.  

$h(g\$, g\$)\rightarrow \$$, 

$h(g\#, g\#)\rightarrow \$$.

\noindent
 GTRS $R_{Q_2}$ consists of the rules

$h(\$, \$)\rightarrow \$$,

$h(\#, \#)\rightarrow \$$,   

$h(f\$, f\$)\rightarrow \$$, 

$h(f\#, f\#)\rightarrow \$$.  

$h(g\$, g\$)\rightarrow \$$, 

$h(g\#, g\#)\rightarrow \$$.

\noindent Clearly,

 $p\hspace{-1mm}\downarrow_ {R_{P_2 \cup Q_2}}= q\hspace{-1mm}\downarrow_ {R_{P_2 \cup Q_2}}$. 

\noindent Hence procedure $\textit{PRO4}$ outputs 'yes' and   halts in the second   step.

\hfill $\b$ }
\end{exa}


We now show the correctness of Procedure $\textit{PRO4}$. 
 \begin{lem} \label{2fecske} Assume that   
 $R_{i-1} =R_i=P_{i+1}$ and $NORM_{i-1}\subset NORM_i$  for some $i\geq 2$. 
Then 
for each equation  $l\approx r$ of $E\cup E^{-1}$ 
with  $l\in T_\S(X_{k+m})$, $r \in T_\S(X_k\cup X_{[k+m+1,  k+m+\ell]})$, 
$k, m\geq 0$,  $\ell\geq 1$, and 
for any
$u_1, \ldots,  u_{k+m}\in
 sub( p\hspace{-1mm}\downarrow_{R_i})\cup (sub(R_i)-lhs(R_i))$, 
$R_i$  does not reach $p\hspace{-1mm}\downarrow_{R_i}$
 starting from $l[u_1, \ldots, u_{k+m}]\downarrow_{R_i}$ attached to some context.
\end{lem}
\pr By contradiction. Assume that there is an 
 equation  $l\approx r$ of $E\cup E^{-1}$ 
with  $l\in T_\S(X_{k+m})$, $r \in T_\S(X_k\cup X_{[k+m+1,  k+m+\ell]})$, 
$k, m\geq 0$,  $\ell\geq 1$, and 
there are 
$u_1, \ldots,  u_{k+m}\in sub( p\hspace{-1mm}\downarrow_{R_i})\cup (sub(R_i)-lhs(R_i))$ such that 
$R_i$  reaches $p\hspace{-1mm}\downarrow_{R_i}$ starting from $l[u_1, \ldots, u_{k+m}]\downarrow_{R_i}$ attached to some context.
By   $R_i=P_{i+1}$,  we do not put equations in $P_{i+1}$ in item (b) of its definition. Consequently, 
for any 
$v_{k+m+1}$, $\ldots$,  $v_{k+m+\ell}\in NORM_i$,
$$l[u_1, \ldots, u_{k+m}]\hspace{-1mm}\downarrow_{R_i}=  
r[u_1, \ldots, u_m,v_{k+m+1}, \ldots, v_{k+m+\ell} ]\hspace{-1mm}\downarrow_{R_i} \, . $$
Hence by our indirect assumption, 
$R_i$ reaches $p\hspace{-1mm}\downarrow_{R_i}$ starting from 

\noindent 
$r[u_1, \ldots, u_m,v_{k+m+1}, \ldots, v_{k+m+\ell} ]\hspace{-1mm}\downarrow_{R_i} $ attached to some context. 
Hence there is a $u\in C_\S$ 
such that $$u[ r[u_1, \ldots, u_m,v_{k+m+1}, \ldots, v_{k+m+\ell} ]\hspace{-1mm}\downarrow_{R_i} ]\tred R 
p\hspace{-1mm}\downarrow_{R_i}\,. $$
Then 
$$u[ r[u_1, \ldots, u_m,v_{k+m+1}\hspace{-1mm}\downarrow_{R_i}, v_{k+m+2}, \ldots, v_{k+m+\ell} ] ]\tred {R_i}
u[ r[u_1, \ldots, u_m,v_{k+m+1}, \ldots, v_{k+m+\ell} ]\hspace{-1mm}\downarrow_{R_i} ]\tred {R_i} 
p\hspace{-1mm}\downarrow_{R_i}\,. $$
By Lemma \ref{alakit}, 
$v_{k+m+1}\hspace{-1mm}\downarrow_{R_i}\in sub(p\hspace{-1mm}\downarrow_{R_i})\cup 
 (sub(R_i)-lhs(R_i))$. Since $R_{i-1}=R_i$,

$v_{k+m+1}\hspace{-1mm}\downarrow_{R_{i-1}}\in sub(p\hspace{-1mm}\downarrow_{R_{i-1}})\cup 
 (sub(R_{i-1})-lhs(R_{i-1}))\subseteq NORM_{i-1}$.

\noindent 
By definition,  $v_{k+m+1}$ is an arbitrary element of $ NORM_i$. Consequently,
we have $NORM_{i}\subseteq NORM_{i-1}$. This is  a contradiction.

\hfill $\b$


\begin{lem}\label{koros}
Let $i\geq 2$. If $R_{i-1}=R_i=R_{i+1}$ and  $NORM_{i-1}=NORM_i$, then $NORM_i=NORM_{i+1}$. 
\end{lem}
\pr
First we show that $NORM_i \subseteq NORM_{i+1}$.
  Let $s\in NORM_{i}$ be arbitrary. If $s \in sub(p\hspace{-1mm}\downarrow_{R_i} ) \cup (sub(R_{i})-lhs(R_{i})) \cup 
\{\, t\hspace{-1mm}\downarrow_{R_{i}} \mid t\in  NORM_{i-1} \, \}$, then 
$s \in sub(p\hspace{-1mm}\downarrow_{R_{i+1}} ) \cup 
(sub(R_{i+1})-lhs(R_{i+1})) \cup 
\{\, t\hspace{-1mm}\downarrow_{R_{i+1}} \mid t\in  NORM_{i} \, \}$. 
Hence $t\in NORM_{i+1}$. 
If 
 $s= f(t_1, \ldots, t_m)\hspace{-1mm}\downarrow_{R_{i}}$ for some $f\in \S_m$ and $t_1, \ldots, t_m \in NORM_{i-1}$, then
 $s= f(t_1, \ldots, t_m)\hspace{-1mm}\downarrow_{R_{i+1}}$ with  $f\in \S_m$ and $t_1, \ldots, t_m \in NORM_{i}$.
Hence $t \in NORM_{i+1}$.

We now show that 
 $NORM_{i+1} \subseteq NORM_i$. 
Let $s\in NORM_{i+1}$ be arbitrary. If $s \in  sub(p\hspace{-1mm}\downarrow_{R_{i+1}} ) \cup     (sub(R_{i+1})-lhs(R_{i+1})) \cup 
\{\, t\hspace{-1mm}\downarrow_{R_{i+1}} \mid t\in  NORM_{i} \, \}$, then 
$s \in  sub(p\hspace{-1mm}\downarrow_{R_i} ) \cup     (sub(R_{i})-lhs(R_{i})) \cup 
\{\, t\hspace{-1mm}\downarrow_{R_{i}} \mid t\in  NORM_{i-1} \, \}$. 
Hence $t\in NORM_{i}$. 
If $s= f(t_1, \ldots, t_m)\hspace{-1mm}\downarrow_{R_{i+1}}$ for some $f\in \S_m$ and $t_1, \ldots, t_m \in NORM_{i}$, then
 $s= f(t_1, \ldots, t_m)\hspace{-1mm}\downarrow_{R_{i}}$ for $f\in \S_m$ and $t_1, \ldots, t_m \in NORM_{i-1}$.
Hence $t \in NORM_{i}$. 

\hfill $\b$


 \begin{lem} \label{2szalag} For each  $i\geq 2$, if 
 $R_{i-1}=R_i=P_{i+1}$, then 
 $R_i=R_{i+1}=P_{i+2}$. 
\end{lem}
\pr By the assumption $R_i=P_{i+1}$ and the definition of $R_{i+1}$, we have 
\begin{equation}\label{buda}
R_i=R_{i+1}\, .
\end{equation}
We now distinguish two cases.

{\em Case 1:} $NORM_{i-1}= NORM_{i}$. By Lemma \ref{koros}, 
\begin{equation}\label{szepen}
NORM_{i}= NORM_{i+1}\,. 
\end{equation}
By (\ref{buda}) and (\ref{szepen}), $P_{i+1}=P_{i+2}$. By the assumption $R_i=P_{i+1}$ and  (\ref{buda}), we have
$R_i=R_{i+1}=P_{i+2}$. 

{\em Case 2:} $NORM_{i-1}\subset NORM_{i}$. Then by
Lemma \ref{2fecske}, 
for each equation  $l\approx r$ of $E\cup E^{-1}$ 
with  $l\in T_\S(X_{k+m})$, $r \in T_\S(X_k\cup X_{[k+m+1,  k+m+\ell]})$, 
$k, m\geq 0$,  $\ell\geq 1$, and 
for any
$u_1, \ldots,  u_{k+m}\in (sub(R_i)-lhs(R_i))\cup sub( p\hspace{-1mm}\downarrow_{R_i})$, 
$R_i$  does not reach $p\hspace{-1mm}\downarrow_{R_i}$
 starting from $l[u_1, \ldots, u_{k+m}]\downarrow_{R_i}$ attached to some context.
Then by (\ref{buda}), 
we do not put equations in $P_{i+2}$ in item (b) in the definition  of  $P_{i+2}$.
Hence $R_{i+1}=P_{i+2}$. By (\ref{buda}) the proof is complete. 

\hfill $\b$

\noindent
Lemma \ref{2szalag} implies the following.
 \begin{lem} \label{2tanc} For each   $i\geq 1$, if 
 $R_{i-1}=R_i=P_{i+1}$,   then for each $k\geq 1$, 
 $R_i=R_{i+k}=P_{i+k+1}$. 
\end{lem}

We now show the correctness of Procedure $\textit{PRO4}$. 
\begin{lem}\label{2romaP}
For any
$n\geq 1$, $t_1, \ldots, t_n \in \ts$, if
$p\thue E t_1 \thue E t_2 \thue E \cdots \thue E t_n$,
then there is $i\geq 1$ such that 
$p\tthue {P_i} t_1 \tthue  {P_i} t_2 \tthue  {P_i} \cdots \tthue  {P_i} t_n$.
\end{lem}
\pr We proceed by induction on $n$.

{\em Base Case}: $n=1$.  Assume that 
$p\thue E t_1$.  
Then   there is an equation
 $l\approx r$ of  $E\cup E^{-1}$ with $l\in T_\S(X_{k+m})$, $r \in T_\S(X_{k+m+\ell})$,
$k, m, \ell\geq 0$, 
and there is  $u\in C_\S$, 
$u_1, \ldots, u_{k+m}, v_{k+m+1}, \ldots, v_{k+m+\ell}\in\ts$ such that 
\begin{equation}\label{02kacsa}
p=u[l[u_1, \ldots, u_{k+m}]] 
\end{equation}  and
$t_{1}=u[r[u_1, \ldots, u_k, v_{k+m+1}, \ldots, v_{k+m+\ell}]]$. 
Let $i=max\{\, height(v_{k+1}), \ldots, height(v_{k+m+\ell})\, \}$. 
By Lemma \ref{talalkozas}, 
$v_{k+m+1}\hspace{-1mm}\downarrow_{R_i}, \ldots, v_{k+m+\ell}\hspace{-1mm}\downarrow_{R_i}$
 are in 
$NORM_i$. 
 By (\ref{02kacsa}), 
 $R_i$  reaches $p\hspace{-1mm}\downarrow_{R_i}$ from $l[u_1\hspace{-1mm}\downarrow_{R_i}, \ldots, u_{k+m}\hspace{-1mm}\downarrow_{R_i}]\hspace{-1mm}\downarrow_{R_i}$ attached to some context. By the definition of $P_{i+1}$, the equation
$$ l[u_1\hspace{-1mm}\downarrow_{R_i}, \ldots, u_{k+m}\hspace{-1mm}\downarrow_{R_i}] \approx r[u_1\hspace{-1mm}\downarrow_{R_i}, \ldots, u_k\hspace{-1mm}\downarrow_{R_i}, 
v_{k+m+1}\hspace{-1mm}\downarrow_{R_i}, \ldots, v_{k+m+\ell}\hspace{-1mm}\downarrow_{R_i}]$$
 is in $\tthue {P_i}$ or  $P_{i+1}$. 
Hence, by the definition of $R_i$ and Statement \ref{2babyP}, 

$p=u[l[u_1, \ldots, u_{k+m}]] \tthue {P_{i+1}} u[l[u_1\hspace{-1mm}\downarrow_{R_i} , \ldots, u_{k+m}\hspace{-1mm}\downarrow_{R_i} ]] 
\tthue {P_{i+1}}$ 

$u[r[u_1\hspace{-1mm}\downarrow_{R_i} , \ldots, u_k\hspace{-1mm}\downarrow_{R_i}, v_{k+m+1}\hspace{-1mm}\downarrow_{R_i}, 
\ldots, v_{k+m+\ell}]\hspace{-1mm}\downarrow_{R_i} ]
\tthue {P_i}
u[r[u_1, \ldots, u_k, v_{k+m+1}, \ldots, v_{k+m+\ell}]]=t_{1}$.

\noindent
Then we have 
$p\tthue {P_{i+1}} t_1$. 

{\em Induction Step}: Let $n\geq 1$, and assume that the satement holds for $1, 2, \ldots, n$.
We now show that the  satement holds for $n+1$.
To this end, assume that 
\begin{equation}\label{kosar2}
p\thue E t_1 \thue E t_2 \thue E \cdots \thue E t_{n+1} \,.
\end{equation}
By the induction  hypothesis, there is $j \geq 1$ such that 
\begin{equation}\label{oras}
p\tthue {P_j} t_1 \tthue  {P_j} t_2 \tthue  {P_j} \cdots \tthue  {P_j} t_{n}\, .
\end{equation}
Hence 
\begin{equation}\label{2gg}
t_n \tred {R_i} p\hspace{-1mm} \downarrow_{R_i} \, . 
\end{equation}
By (\ref{kosar2}),  there is an equation
 $l\approx r$ in $E\cup E^{-1} $ with 
 $l\in T_\S(X_{k+m})$, $r \in T_\S(X_k\cup X_{[k+m+1,  k+m+\ell]})$ for some 
$k, m, \ell\geq 0$, 
and there are   $u\in C_\S$, 
$u_1, \ldots, u_{k+m}, v_{k+m+1}, \ldots, v_{k+m+\ell}\in\ts$ such that 
\begin{equation}\label{kacsa}
t_n=u[l[u_1, \ldots, u_{k+m}]] \mbox{  and }
t_{n+1}=u[r[u_1, \ldots, u_k, v_{k+m+1}, \ldots, v_{k+m+\ell}]]\, .
\end{equation}
Let $i=max\{\, j, height(v_{k+m+1}), \ldots, height(v_{k+m+\ell})\, \}$. 
By Lemma \ref{talalkozas}, 
 $v_{k+m+1}\hspace{-1mm}\downarrow_{R_i}, \ldots, v_{k+m+\ell}\hspace{-1mm}\downarrow_{R_i}$
 are in 
$NORM_i$.  
 Clearly, 
$l[u_1\hspace{-1mm}\downarrow_{R_i}, \ldots, u_{k+m}\hspace{-1mm}\downarrow_{R_i}] \tred {R_i} 
l[u_1\hspace{-1mm}\downarrow_{R_i}, \ldots, u_{k+m}\hspace{-1mm}\downarrow_{R_i}]\hspace{-1mm}\downarrow_{R_i}
$.
Then by  (\ref{2gg}) and (\ref{kacsa}),
 $R_i$  reaches $p\hspace{-1mm}\downarrow_{R_i}$ starting from $l[u_1\hspace{-1mm}\downarrow_{R_i}, \ldots, u_{k+m}\hspace{-1mm}\downarrow_{R_i}]\hspace{-1mm}\downarrow_{R_i}$ attached to some context. By the definition of $P_{i+1}$, the equation

$ l[u_1\hspace{-1mm}\downarrow_{R_i}, \ldots, u_{k+m}\hspace{-1mm}\downarrow_{R_i}] \approx r[u_1\hspace{-1mm}\downarrow_{R_i}, \ldots, u_k\hspace{-1mm}\downarrow_{R_i}, 
v_{k+m+1}\hspace{-1mm}\downarrow_{R_i}, \ldots, v_{k+m+\ell}\hspace{-1mm}\downarrow_{R_i}]$

\noindent is in $\tthue {P_i}$ or $P_{i+1}$. 
Hence, by the definition of $R_i$  and Statement \ref{2babyP}, 

$t_n=u[l[u_1, \ldots, u_{k+m}]] \tthue {P_{i+1}} u[l[u_1\hspace{-1mm}\downarrow_{R_i}, \ldots, u_{k+m}\hspace{-1mm}\downarrow_{R_i}]] 
\thue {P_{i+1}}$ 

$u[r[u_1\hspace{-1mm}\downarrow_{R_i} , \ldots, u_k\hspace{-1mm}\downarrow_{R_i}, v_{k+m+1}\hspace{-1mm}\downarrow_{R_i}, \ldots, v_{k+m+\ell}\hspace{-1mm}\downarrow_{R_i}]]
\tthue {P_{i+1}}
u[r[u_1, \ldots, u_k, v_{k+m+1}, \ldots, v_{k+m+\ell}]]=t_{n+1}$.

\noindent  
By (\ref{oras}),
$p\tthue {P_{i+1}} t_1 \tthue  {P_{i+1}} t_2 \tthue  {P_{i+1}} \cdots \tthue  {P_{i+1}} t_{n}
\tthue  {P_{i+1}} t_{n+1}$.

\hfill $\b$

By Statement \ref{2babyP}, Lemma \ref{2tanc}, and Lemma \ref{2romaP} we have the following result.
\begin{lem}\label{2oszP} For each   $i\geq 2$, if 
 $R_{i-1}=R_i=P_{i+1}$, then
for each $q'\in \ts$, 
$p \tthue {P_i} q'$ if and only if $p \tthue E q'$. 

\end{lem}

We can show the following in the same way as Theorem \ref{egy}.
\begin {tet}\label{one} If $p\tthue E q$, then procedure $\textit{PRO4}$  outputs 'yes' and  halts. 
\end{tet}
We can show the following in the same way as Theorem \ref{ketto}.
\begin {tet} \label{two} If  procedure $\textit{PRO4}$  outputs 'yes'  and  halts, then 
$p \tthue E q$. If procedure $\textit{PRO4}$  outputs 'no' and  halts,  then 
$(p , q) \not \in \tthue E $.

\end{tet}

Theorems \ref{one} and \ref{two} imply the following.
\begin {tet} \label{fo2} If   
$p \tthue E q$, then procedure $\textit{PRO4}$  outputs 'yes'  and  halts. Otherwise,  either $\textit{PRO4}$  outputs 'no' and  halts, or 
 $\textit{PRO4}$ does not halt at all. 

\end{tet}

\section{Comparison with  the Knuth-Bendix completion procedure}\label{foglal}
 We now 
compare  procedures $\textit{PRO3}$ and $\textit{PRO4}$ with the basic  Knuth-Bendix completion procedure 
(see Section 7.1 in \cite{baanip}),  the improved version of the  Knuth-Bendix completion procedure
 described by a set of inference rules 
(see Section 7.2  in \cite{baanip}),
the  goal-directed completion procedure based on SOUR graphs  \cite{lyn,ls}, and the 
unfailing  Knuth-Bendix completion procedure \cite{bdp}. 
In contrast to all  versions of the Knuth-Bendix procedure, 
Procedures $\textit{PRO3}$ and $\textit{PRO4}$ do not compute any critical pairs and do not use a  reduction order. 
They do not attempt to construct a convergent TRS equivalent to $E$.
When $\textit{PRO3}$ and $\textit{PRO4}$ run a congruence closure algorithm for the TES $E$ over the subterm graph of $E$ \cite{dst,nelopp}, 
they compute and then process only finitely many 
ground instances $( \overline{s}, \overline{t} )$ of  finitely many elements   
$(s, t)$ of the relation $\tthue E$, where $s, t$ may contain variables.
  Here $(s, t)$ need not be a critical pair computed by the basic  Knuth-Bendix completion procedure.
 In fact, the ground instances
$( \overline{s}, \overline{t})$ are elements of the equivalence relation 
$\tthue E \cap (sub(E)\times sub(E))$.
Procedures $\textit{PRO3}$ and $\textit{PRO4}$
compute a representative  $r$  of $\overline{s}$ and  $\overline{t}$ for the 
equivalence relation  $\tthue E \cap (sub(E)\times sub(E))$. 
The representative $r$ becomes the normal form of $\overline{s}$ and  $\overline{t}$ for the rewrite relation induced by the constructed reduced GTRS.
Hence,  $\textit{PRO3}$ and $\textit{PRO4}$ do not  compare  the normal forms of $s$ and  $t$
via any reduction order. 
 In contrast, the basic  Knuth-Bendix completion procedure reduces the terms 
in each critical pair to their normal forms. Then 
tries to orient the  normal forms into a  rewrite rule.   
In this way the procedure orients all instances of these terms as well.   
The improved version of the  Knuth-Bendix completion procedure
 described by a set of inference rules 
(see Section 7.2  in \cite{baanip})
also processes each  critical pair and 
also orients the obtained pair, and hence all of its  instances. 
The unfailing  Knuth-Bendix completion procedure \cite{bdp}
 applies orientable instances of equations in $E$ with respect to a reduction order $>$.

To illustrate the efficiency of the  goal-directed completion procedure, 
Lynch \cite{lyn} presented the following example. Let the ranked alphabet $\S$ consist of the unary symbols $f$, $g$
 and the nullary symbols
$\$$, $\#$. 
Consider the variable preserving TES $E=\{\, ffx\approx gfx\, \}$. We raise the problem whether
$\$\tthue E \#$. On the one hand,  the basic Knuth-Bendix completion procedure runs forever on this example  \cite{lyn}. 
On the other hand,  
the goal-directed completion  procedure
does not generate any rule applicable to $\$$ or  $\#$. 
Therefore, the goal-directed completion procedure outputs
'no' and halts \cite{lyn}. 
In the terminology of  Lynch and Strogova \cite{ls}, we say  that
 ``the goal-directed completion  procedure
compiles the TES $E$ and the goal $(p, q)$. After the compilation is finished, 
we cannot apply a schematization of an equation in the completed system.
Therefore, the goal-directed completion procedure outputs
'no' and halts. 
This is an example
where the goal-directed completion  procedure is superior to the basic Knuth-Bendix algorithm.'' \cite{ls}
It is still open whether  the goal-directed completion procedure halts on the TES $E$ and any goal  \cite{lyn}.
As for the above example, $\textit{PRO3}$  gives the correct answer and then halts on the TES $E$  and any terms $p, q\in \ts$.

We conjecture that there are variable preserving TES $E$  and ground terms $p, q$ such that Conditions (a)-(c) hold.

(a) 
The basic Knuth-Bendix completion procedure runs forever on $E$.

(b)
There is a goal $(p, q)$ such that the goal-directed completion procedure does not stop  on  $E$ and  $(p, q)$.

(c) Procedure $\textit{PRO3}$  gives the correct answer and then halts on the TES $E$  and any terms $p, q\in \ts$.

\noindent 
Let TES $E$ be as in Example \ref{dalol}.
We conjecture that  there is $q\in \ts$ such that the   symbols  $a, c$ do not appear in $q$
and that 
the goal-directed completion procedure does not halt on the TES $E$ and the goal  $(aaa\$, q)$.
On the other hand, let  $q\in \ts$ be arbitrary such that the
symbols  $a, c$ do not appear in $q$.    On the input $E$,  $aaa\$$, $q$, Procedure 
$\textit{PRO3}$ outputs 'no', the correct answer, and then  halts, see Example \ref{dalol}.

Procedures $\textit{PRO3}$ and  $\textit{PRO4}$ attempt to construct the reduced   
GTRSs $R_P$ and  $R_Q$, rather than a convergent term rewrite system equivalent to $E$,
such that 

$\bullet$ $R_P \cup R_Q \subseteq \tthue E$,
 
$\bullet$ $p\tthue {R_P} q$ or 
$\tthue {R_P}\cap (\{\,  p\, \}\times \ts)=\tthue E\cap (\{\,  p\, \}\times \ts) $, and 

$\bullet$ $p\tthue {R_Q} q$ or $\tthue {R_Q}\cap (\{\,  q\, \}\times \ts)=\tthue E\cap (\{\,  q\, \}\times \ts) $. 

\noindent
Thus $R_P$ and  $R_Q$ need not  be equivalent to $E$. 
By contrast,  all versions of the Knuth-Bendix   completion procedure attempt to transform a given TES $E$ into an equivalent 
convergent term rewrite sytem.
Since Snyder's  ground  completion algorithm does not apply orderings, 
procedures $\textit{PRO3}$ and $\textit{PRO4}$ do not apply any orderings as well.

We now present three examples where procedures $\textit{PRO3}$ and  $\textit{PRO4}$ compute efficiently, 
probably more efficiently than all  versions of the Knuth-Bendix completion procedure. 
\begin{exa}{\em \cite{gnprs,plasat} Gallier et al \cite{gnprs} and 
Plaisted and Sattler-Klein \cite{plasat} presented the following problem to illustrate that reducing a ground 
term to its normal form  can take exponential time if a proper strategy is not used.
Let $\S=\S_0\cup \S_1$, $\S_0=\{\, \$\, \}$, and $\S_1=\{\, f, g\, \}$. Let $n\geq 2$.
Let the   GTRS $R$ 
  consist  of the following rules:

$f\$ \rightarrow g\$$, 

$fg\$ \rightarrow gf\$$,

$fg^2\$ \rightarrow gf^2\$$,

$\ldots$

$fg^n\$ \rightarrow gf^n\$$.

\noindent  Plaisted and Sattler-Klein observed the following  on page 156 in \cite{plasat}.
Although GTRS $R$ is convergent, the right-hand sides can be further rewritten. 
An unskilful choice of rewrites can lead to an exponential time of process.
The straightforward reduction of the term
$gf^n\$$ can take a number of rewrite steps exponential in $n$. 
However, if we apply the rules in order of size, smallest first, to all other rules, 
the whole TRS can be rewritten to a reduced GTRS in a polynomial number of steps.

We form the TES $E$ by adding the equation 

$$fg^{n+1}x\approx gf^{n+1}x$$

\noindent to the set $R$. 
We now run procedure $\textit{PRO3}$
 on the variable preserving   TES $E$ and the ground terms  $p=f^{n+2}\$$ and $q=g^{n+2}\$$.
Then 

$\{\, f\$ \approx  g\$\, \}=P_1=R_{P_1}=Q_1= R_{Q_1} $, $p\hspace{-1mm}\downarrow_{R_{P_{1}\cup Q_{1}}}\neq 
q\hspace{-1mm}\downarrow_{R_{P_{1}\cup Q_{1}}}=q$.

$R_1\cup \{\, fg\$ \approx  g^2\$\, \}=P_2=R_{P_2}=Q_2= R_{Q_2} $, $p\hspace{-1mm}\downarrow_{R_{P_{2}\cup Q_{2}}}\neq 
q\hspace{-1mm}\downarrow_{R_{P_{2}\cup Q_{2}}}=q$.

$R_2\cup\{\, fg^2\$ \approx  g^3\$\, \}=P_3=R_{P_3}=Q_3= R_{Q_3} $, 
 $p\hspace{-1mm}\downarrow_{R_{P_{3}\cup Q_{3}}}\neq 
q\hspace{-1mm}\downarrow_{R_{P_{3}\cup Q_{3}}}=q$.

$\ldots$

$R_{P_{n}}\cup \{\, fg^n\$ \approx  g^n\$\, \}=P_n=R_{P_n}=Q_n= R_{Q_n} $, 
$p\hspace{-1mm}\downarrow_{R_{P_{n+1}\cup Q_{n+1}}}\neq
q\hspace{-1mm}\downarrow_{R_{P_{n+1}\cup Q_{n+1}}}$.

$R_{P_{n+1}}\cup \{\, fg^{n+1}\$ \approx  g^{n+1}\$\, \}=P_{n+2}=R_{P_{n+2}}=Q_{n+2}= R_{Q_{n+2}} $. 

$P_{n+2}=R_{P_{n+3}}=Q_{n+2}=R_{Q_{n+3}}  $.

\noindent 
Observe that 
$p\hspace{-1mm}\downarrow_{R_{P_{n+2}\cup Q_{n+2}}}= 
q\hspace{-1mm}\downarrow_{R_{P_{n+2}\cup Q_{n+2}}}=q$.
 Hence procedure
  $\textit{PRO3}$    outputs 'yes' and  halts in the $(n+2)$nd step. 
The number of computation steps is polynomial.  
  It should be clear that for all ground terms $p$ and $q$,  $\textit{PRO3}$ halts. It  outputs 'yes'   if 
$p \tthue E q$. Otherwise it  outputs 'no'.

Consider the lexicographic path order  $>_{lpo}$ induced by the order $f > g > \$$ \cite{baanip}. 
We now run the basic Knuth-Bendix completion procedure on the TES $E$ and the reduction order $>_{lpo}$.
In the initialization
phase, the basic Knuth-Bendix completion procedure orients the equations of $E$.
We obtain the TRS $S$ consisting of the following rules:

$f\$ \rightarrow g\$$, 

$fg\$ \rightarrow gf\$$,

$fg^2\$ \rightarrow gf^2\$$, 

$\ldots$

$fg^n\$ \rightarrow gf^n\$$, 

$fg^{n+1}x \rightarrow  gf^{n+1}x$. 

\noindent Similarly to the first part of the example we have the following.
The TRS $S$ has no critical pairs. Hence the basic Knuth-Bendix procedure outputs $S$.
The straightforward reduction of the term
$f^{n+2}\$$ to $g^{n+2}\$$ by $S$ takes a number of rewrite steps exponential in $n$. 
The improved Knuth-Bendix completion procedure reduces the right-hand sides of the first  $n$ rules
as in the first part of the example. 
We obtain the TRS $S'$ consisting of the following rules:

$f\$ \rightarrow g\$$, 

$fg\$ \rightarrow gf\$$, 
$fg\$ \rightarrow gg\$$, 

$fg^2\$ \rightarrow gf^2\$$, $fg^2\$ \rightarrow gfg\$$, $fg^2\$ \rightarrow g^3\$$, 

$\ldots$

$fg^n\$ \rightarrow gf^n\$$, $fg^n\$ \rightarrow gf^{n-1}g\$$, $\ldots$, $fg^n\$ \rightarrow g^{n+1}\$$,

$fg^{n+1}x \rightarrow  gf^{n+1}x$. 

\noindent
In the best case,   the  reduction of the term
$f^{n+2}\$$ to $g^{n+2}\$$ applies the rules 

$f\$ \rightarrow g\$$,

$fg\$ \rightarrow gg\$$,  

$fg^2\$ \rightarrow g^3\$$,

$\ldots$

 $fg^n\$ \rightarrow g^{n+1}\$$,

$fg^{n+1}x \rightarrow  gf^{n+1}x$.

\noindent
In the worst  case, $S'$ applies only the rules of $S$
 in the  reduction of the term
$f^{n+2}\$$ to $g^{n+2}\$$. 
 Hence it takes a number of rewrite steps exponential in $n$ as in the first part of the example.
The goal-directed completion procedure  computes fast on $E$ and the goal $(p, q)$. For  experimental results,
 see the line of the problem Counter5 in
Table 1 in Section 7 in \cite{ls}. 

\hfill $\b$
}\end{exa}

\begin{exa}{\em \cite{gnprs,plasat} We modify an example of 
Plaisted and Sattler-Klein \cite{plasat} and Lynch and Strogova \cite{ls}. Let $n\geq 2$.
Let $\S=\S_0\cup \S_2$, $\S_0=\{\, \$_1, \$_2, \ldots, \$n, \#_1, \#_2, \ldots, \#_n \, \}$, and $\S_2=\{\, f, g\, \}$. 
Let the   TES $E$ 
  consist  of the following equations:

$ f(\flat, \flat)\approx f(\#_0,  \$_0 )$, 

$\$_0 \approx f(\$_1, \#_1)$, 

$\#_0 \approx g(\#_1,\$_1)$, 

$\$_1 \approx f(\$_2, \#_2)$, 

$\#_1 \approx g(\#_2,\$_2)$, 

$\ldots$

$\$_{n-1} \approx f(\$_n, \#_n)$, 

$\#_{n-1} \approx g(\#_n,\$_n)$,  

 $\$_n  \approx \#_n$,

$f(x_1, x_1) \approx g(x_1, x_1)$.

We now run procedure $\textit{PRO3}$
 on the variable preserving   TES $E$ and the ground terms  $p=f(\$_0, \#_0)$ and $q=g(\#_0, \#_0)$.
Then 

$\{\,  f(\$_1, \#_1)\approx \$_0,\,   g(\#_1,\$_1) \approx  \#_0  \, \}=P_1=R_{P_1}$,

$\{\, g(\#_1,\$_1) \approx  \#_0, \, f(\#_0, \#_0)\approx g(\#_0, \#_0) 
\, \}=Q_1= R_{Q_1} $,

$R_{P_1}\cup \{\,  f(\$_2, \#_2)\approx \$_1, \,  g(\#_2,\$_2) \approx  \$_1
\, \}=P_2=R_{P_2}$,

$ R_{Q_1}\cup \{\,   f(\$_1, \#_1)\approx \$_0, \,  f(\$_2, \#_2)\approx \$_1, \,  g(\#_2,\$_2) \approx\#_1 \, \}=Q_2= R_{Q_2} $,

$\ldots$

$R_{P_{n-1}}\cup \{\, f(\$_n, \#_n) \approx  \$_{n-1} ,\,  g(\#_n,\$_n) \approx\#_{n-1}
\, \}=P_n=R_{P_n}$,

$ R_{Q_{n-1}}\cup \{\,   f(\$_{n-1}, \#_{n-1})\approx  \$_{n-2} ,\,  g(\#_n,\$_n) \approx\#_{n-1}
 \, \}=Q_n= R_{Q_n} $.

\noindent
$R_{P_{n}}$ consists of the following rules:

$f(\$_1, \#_1)\rightarrow \$_0$,  

$ g(\#_1,\$_1) \rightarrow  \#_0$,

$f(\$_2, \#_2)\rightarrow \$_1$, 

$ g(\#_2,\$_2) \rightarrow  \#_1$,

$\ldots$

$f(\$_n, \#_n) \rightarrow  \$_{n-1} $, 

$g(\#_n,\$_n) \rightarrow\#_{n-1}$,


$R_{P_{n}}\cup \{\, \$_n \approx  \#_n\, \}=P_{n+1}$.

\noindent
$R_{P_{n+1}}$ consists of the following rules:

$f(\$_0, \#_0) \rightarrow \$_0$,

$f(\$_1, \$_1)\rightarrow \$_0$,  

$f(\$_2, \$_2)\rightarrow \$_1$,

$\ldots$

$g(\$_n,\$_n) \rightarrow\$_{n-1}$,

$\#_0 \rightarrow \$_0$,

$\#_1 \rightarrow \$_1$,

$\#_2 \rightarrow \$_2$,

$\ldots$

$\#_n \rightarrow \$_n$.

$R_{Q_{n}}\cup \{\,  f(\$_n, \#_n) \approx \$_{n-1},\,  \$ \approx  \#\, \}=Q_{n+1}$,

$R_{Q_{n+1}}=R_{P_{n+1}}\cup \{\, f(\#_0, \#_0)\rightarrow g(\#_0, \#_0)
 \,\}$.

$P_{n+2}=R_{P_{n+3}}=Q_{n+2}=R_{Q_{n+3}}  $. 

\noindent
Clearly, $p\hspace{-1mm}\downarrow_{R_{P_{n+1}}}=q\hspace{-1mm}\downarrow_{R_{P_{n+1}}}$.
Consequently, procedure
  $\textit{PRO3}$    outputs 'yes' and  halts in the $(n+1)$st step. 
The number of computation steps is polynomial.  

Consider the lexicographic path order  $>_{lpo}$ induced by the order  $\flat >\$_0> \$_1>\cdots \$_n>
\#_0> \#_1>\cdots \#_n  > f >g $. 
  We now run the basic Knuth-Bendix completion procedure on the TES $E$ and the reduction order $>_{lpo}$.
In the initialization
phase, the basic Knuth-Bendix completion procedure orients the equations of $E$.
We obtain the TRS $S$ consisting of the following rules:

$\$_0 \rightarrow f(\$_1, \#_1)$, 

$\#_0 \rightarrow g(\#_1,\$_1)$, 

$\$_1 \rightarrow f(\$_2, \#_2)$, 

$\#_1 \rightarrow g(\#_2,\$_2)$, 

$\ldots$

$\$_{n-1} \rightarrow f(\$_n, \#_n)$, 

$\#_{n-1} \rightarrow g(\#_n,\$_n)$,  

 $\$_n  \rightarrow \#_n$, 

$f(\flat, \flat) \rightarrow f(\#_0, \$_0)$,

$f(x_1, x_1) \rightarrow g(x_1, x_1)$.

\noindent The  last two  rules yield the 
critical pair $\langle  f(\#_0, \$_0), g(\flat, \flat)\rangle$. 
Observe that $f(\#_0, \$_0)$ has a unique $\red S$ normal form, and that 
$size(f(\#_0, \$_0)\hspace{-1mm}\downarrow_S)=2^{n+1}$. 
Thus the completed system contains a rule with a left-hand side of size $2^{n+1}$. 
The improved Knuth-Bendix completion procedure  also yields the TRS $S$ and the above critical pair. 
Again, the completed system contains a rule with a left-hand side of size $2^{n+1}$. 
The goal-directed completion procedure based on SOUR graphs \cite{lyn,ls} stores the term  
$f(\#_0, \$_0)\hspace{-1mm}\downarrow_S$ 
 in linear space in  $n$.

 \hfill $\b$
}\end{exa}

\begin{exa}{\em Let $\S=\S_0\cup \S_1$, $\S_0=\{\, \$ \, \}$, and $\S_1=\{\, a, b\, \}$. 
Let the   GTES $F$ 
  consist  of the equation $abbax_1 \approx x_1$. 
Furthermore, let the   GTES $E$ 
  consist  of the equations 

$abbax_1 \approx x_1$,   $a\$ \approx \$$,  $b\$ \approx \$$.

\noindent It is well-known that there is no convergent TRS $R$ equivalent to $F$, see Theorem  4.2.18 in \cite{jan}. 
Hence there is no convergent TRS $R$ equivalent to $E$ either. Consequently, 
 the basic  Knuth-Bendix completion procedure 
(see Section 7.1 in \cite{baanip}), the improved version of the  Knuth-Bendix completion procedure
 described by a set of inference rules 
(see Section 7.2  in \cite{baanip}) cannot produce a convergent TRS $R$ equivalent to $E$.

Let $p, q\in \ts$ be arbitrary. First,
we run the procedure  $\textit{PRO3}$ on the input
$E$,  $p$, $q$. Procedure 
  $\textit{PRO3}$    outputs 'yes' and  halts in the first or second step. The resulting reduced GTRS is a subset of

$\{\, a\$ \rightarrow \$, \,  b\$ \rightarrow \$ \, \}$. 

\noindent 
Second, we run 
the  goal-directed completion procedure on the input $E$, $(p, q)$.
It computes all critical pairs and then processes them. Then it 
applies the resulting rules. The  goal-directed completion procedure takes 
more time on $E$ and the goal $(p, q)$ than  procedure $\textit{PRO3}$ on the input
$E$,  $p$, $q$.

\hfill $\b$
}\end{exa}


\section{Conclusion}\label{osszefog}
We recalled the well known trivial  semi-decision procedure $\textit{PRO1}$
 for the ground word problem of  variable preserving TESs and its straightforward generalization, 
the  trivial  semi-decision procedure $\textit{PRO2}$  for the ground word problem of TESs. 
On the basis of $\textit{PRO1}$, we  gave the    semi-decision procedure $\textit{PRO3}$ for the ground
 word problem of   variable preserving TESs. We gave examples when procedure $\textit{PRO3}$ was more efficient  than procedure $\textit{PRO1}$. 
Then we presented the  semi-decision procedure $\textit{PRO4}$
 for the ground word problem of term equation systems. We obtained  it  generalizing  $\textit{PRO3}$ taking into account  $\textit{PRO2}$. 
We showed the  correctness of $\textit{PRO3}$ and $\textit{PRO4}$. 
We compared the  procedures $\textit{PRO3}$ and $\textit{PRO4}$ with the basic Knuth-Bendix completion procedure
and 
the  goal-directed completion procedure based on SOUR graphs  \cite{lyn,ls}.

Procedures   $\textit{PRO3}$ or $\textit{PRO4}$ compute in a different way than all versions of  the 
Knuth-Bendix completion procedure.  
 To some instances of   the ground word problem of a TES $E$,
they give an answer 
sooner than all versions of the  Knuth-Bendix completion procedure
or it is open whether some version  of the  Knuth-Bendix completion procedure gives an answer at all.
Assume that, given a TES $E$ and ground terms $p, q$, we want to decide whether $p\tthue E q$. 
The ground word problem is undecidable even for variable-preserving TESs.
Consequently, we  have no upper bound on the running time of any type of the  Knuth-Bendix completion procedure on the input  
TES $E$ any reduction order $>$ and the ground terms $p, q$. 
 However, we assume beforehand that the basic Knuth-Bendix completion procedure or the goal-directed completion procedure  or
the nonfailing Knuth-Bendix completion procedure
will stop  on $E$, $>$,  and $p, q$, and estimate its running time.  
We base our time estimate on the size of the input and the experimental results  
 by the various implementations \cite{ghls, hil, mec, wsw} of all versions of the  Knuth-Bendix completion procedure 
on inputs of similar size. 
Then we carry out the following steps. Simultaneously, we start all implementations  of all versions
of the Knuth-Bendix completion procedure on $E$ and 
$p, q$. We wait for the  estimated running time.  If none of the procedures  stop within this time,  then they do not stop at all, or we underestimated the  running time.
Then we start the 
 procedure $\textit{PRO3}$ or $\textit{PRO4}$ depending on whether TES $E$ is variable preserving. 
In some cases
 $\textit{PRO3}$ or $\textit{PRO4}$ might give  an  answer sooner than all
 implementations  of all versions of the Knuth-Bendix completion procedure. 

We presented ad hoc examples 
 when procedure $\textit{PRO3}$ was probably more efficient  than the goal-directed completion procedure   \cite{lyn,ls}. However, 
to justify the introduction of  procedures 
 $\textit{PRO3}$ and $\textit{PRO4}$, 
we need further evidence for the efficiency of the procedures 
 $\textit{PRO3}$ and $\textit{PRO4}$. We should present implementation results and theoretical arguments. 
We now raise  questions on the efficiency of $\textit{PRO3}$ and  $\textit{PRO4}$ compared to the  various versions of the Knuth-Bendix completion procedure.
\begin{ques}\label{tztzt} Is it true that 
for most instances of the ground  word problem of a TES $E$,  a correctly chosen version of the 
 Knuth-Bendix completion procedure is more efficient than $\textit{PRO3}$ or  $\textit{PRO4}$? 
\end{ques}

\begin{ques}
For which instances of the ground  word problem of a TES $E$,  is a correctly chosen version of the 
 Knuth-Bendix completion procedure more efficient than $\textit{PRO3}$ or  $\textit{PRO4}$? 
\end{ques}

\begin{ques}
Is it decidable for an  instance
 of the ground  word problem of a TES $E$,  whether
a correctly chosen version of the 
 Knuth-Bendix completion procedure is  more efficient than $\textit{PRO3}$ or  $\textit{PRO4}$? 
\end{ques}

\begin{ques}
Is there  an  instance
 of the ground  word problem of a TES $E$, such that 
no version of the 
 Knuth-Bendix completion procedure halts, and  $\textit{PRO3}$ or  $\textit{PRO4}$ halts? 
\end{ques}

We can reduce an instance of the word problem for a TES $E$  to an instance of the  ground word problem for $E$ over a larger alphabet $\Delta$.
Let $E$ be a TES and $p, q$ arbitrary terms  over a ranked alphabet $\S$. 
Assume that exactly the variables $x_1, \ldots, x_m$ appear in $p$ or $q$. We now define the ranked alphabet $\Delta$. 
It  contains each element of $\S$.
Furthermore, for each $i=1, \ldots, m$, we add a  new constant $\#_i$ to $\Delta$.
We define $p'$ from $p$  and $q'$ from $q$ by replacing each occurrence of $x_i$ with $\#_i$  for $i=1, \ldots, m$.  
Then $p\tthue E q$ over $\S$ if and only if  $p' \tthue E q'$ over $\Delta$. Thus if we can decide whether  $p' \tthue E q'$ over $\Delta$,
 then we can also decide whether $p\tthue E q$  over $\S$.

\end{document}